\documentclass[12pt]{article}
\usepackage{amsmath,bbold,amsfonts,amssymb,graphics,xcolor}
\usepackage[T1]{fontenc}
\usepackage{graphicx}
\usepackage{tikz}
\usetikzlibrary{positioning}
\makeatletter

\numberwithin{equation}{section}

\def\uqsu2{\mathcal{U}_{\mathsf{q}}(\mathfrak{su}_2)}

\newcommand{\beq}{\begin{equation}}
\newcommand{\eeq}{\end{equation}}
\newcommand{\bea}{\begin{eqnarray}}
\newcommand{\eea}{\end{eqnarray}}
\newcommand{\beaq}{\begin{eqnarray}}
\newcommand{\eeaq}{\end{eqnarray}}
\newcommand{\infint}{\int_{-\infty}^\infty}

\begin{document}
\begin{flushright}\footnotesize
\end{flushright}
\vspace{0.7cm}
\centerline{\Large \bf New RG flows between non-Unitary CFTs }
\vskip 0.5cm
\centerline{\Large \bf from exact massless scattering theories}
\vskip 1cm
\centerline{\large Changrim Ahn$^{\rm a}$\,\footnote{ahn@ewha.ac.kr}, 
Zoltan Bajnok$^{\rm b}$\,\footnote{bajnok.zoltan@wigner.hu}, and
Soma Elek$^{\rm b,c}$\,\footnote{elek.soma@wigner.hu}}
\vskip .5cm
\centerline{\it$^{\rm a}$Department of Physics,}
\centerline{\it Ewha Womans University}
\centerline{\it Seoul 120-750, Korea}
\vskip 0.5cm
\centerline{\it$^{\rm b}$HUN-REN Wigner Research Centre for Physics,}
\centerline{\it  Konkoly-Thege Miklós út 29-33, 1121 Budapest, Hungary}
\vskip 0.5cm
\centerline{\it$^{\rm c}$ Roland Eötvös University, Department for Theoretical Physics, }
\centerline{\it 1117 Budapest, Pázmány sétány 1/A, Hungary }
\vskip 1cm
\centerline{\small PACS: 11.25.Hf, 11.55.Ds}
\vskip 1cm

\centerline{\bf Abstract}

We construct complete solutions of the massless S-matrix bootstrap equations describing a new infinite family of integrable RG flows from the higher-fusion-level minimal models ${\cal M}(3,2n+3;n)$ to the non-unitary Virasoro minimal models ${\cal M}(2,2n+3)$, including a flow from ${\cal M}(3,5)$ to the Yang--Lee CFT ${\cal M}(2,5)$ when $n=1$. We provide strong evidence for this identification by matching conformal perturbation theory around the UV fixed points with the small-volume expansion of the corresponding thermodynamic Bethe ansatz equations. We further show that a family of non-invertible Verlinde defect lines is preserved along the flows.

\newpage
\section{Introduction}

Renormalization-group (RG) flows between conformal field theories (CFTs) provide a fundamental organizing principle in quantum field theory. They describe how the effective degrees of freedom and interactions change with the energy scale, interpolating between ultraviolet (UV) and infrared (IR) fixed points. Distinct UV theories may flow to the same IR CFT, thereby revealing universal long-distance behavior that is insensitive to microscopic details.

In two dimensions, conformal symmetry provides exceptionally powerful tools for characterizing the fixed points through their central charges, operator spectra, and correlation functions. RG flows between unitary two-dimensional CFTs have consequently been studied by a wide range of analytical and numerical methods. A central structural result is Zamolodchikov's $c$-theorem, which establishes the existence of a function that decreases monotonically along unitary RG flows and coincides with the central charge at the fixed points \cite{ZamRG,Zam_c}. In particular, a unitary flow from a UV CFT to an IR CFT must satisfy
$c_{\mathrm{UV}}>c_{\mathrm{IR}}$.

The situation is considerably less constrained for non-unitary theories. Reflection positivity is absent, and neither the monotonicity argument underlying the $c$-theorem nor the usual ordering of fixed points by their central charges applies directly. Nevertheless, non-unitary CFTs and the RG flows connecting them arise naturally in statistical mechanics, critical phenomena, and integrable quantum field theory. Understanding such flows therefore offers a setting in which to investigate universality and exact RG dynamics beyond the unitary paradigm.

A standard approach to two-dimensional RG flows is to perturb a conformal field theory by a relevant operator and analyze the resulting theory using conformal perturbation theory or an effective Ginzburg--Landau description. For integrable quantum field theories, the thermodynamic Bethe ansatz (TBA) provides a powerful non-perturbative framework for following the flow between fixed points. Its derivation, however, requires the exact scattering data of the interpolating theory. In many examples these $S$-matrices are not known, so the proposed TBA equations—and hence the associated RG flows—remain conjectural.

A complementary strategy was introduced in \cite{AhnLeC}. One seeks a massless integrable scattering theory that describes an RG flow generated by a relevant perturbation of a UV CFT and terminating at a prescribed IR CFT. Viewed from the infrared, the same trajectory is obtained by deforming the IR fixed point by irrelevant operators. In the undeformed IR CFT, the left- and right-moving sectors are decoupled. The irrelevant deformation couples them and, when integrability is preserved, gives rise to nontrivial factorized scattering amplitudes between left- and right-moving excitations.

The same-chirality amplitudes, $S^{\rm LL}=S^{\rm RR}$,
are inherited from the massive integrable perturbation associated with the IR CFT and determine its infrared dynamics. The information about the ultraviolet completion is instead encoded in the mixed-chirality amplitudes $S^{\rm RL}$ and $S^{\rm LR}$. Depending on their analytic structure, the resulting massless scattering theory may approach a UV conformal fixed point or terminate at a Hagedorn-type singularity. The central problem is therefore to bootstrap the exact left--right scattering amplitudes.

When several particles have the same mass and carry internal quantum numbers, the same-chirality scattering matrices $S^{\rm RR}$ and $S^{\rm LL}$ may be non-diagonal. The mixed-chirality matrix $S^{\rm RL}$ must then satisfy massless Yang--Baxter equations involving $S^{\rm RR}$ and $S^{\rm LL}$. These equations fix its matrix structure, leaving only scalar factors that encode the choice of irrelevant deformation of the IR theory \cite{AhnBaj}. When the particle masses are non-degenerate, the scattering is diagonal in particle space. The individual amplitudes $S_{ab}^{\rm RL}$ are then not constrained by a nontrivial matrix Yang--Baxter equation, but they remain subject to unitarity, crossing symmetry, analyticity, and the bootstrap relations inherited from the bound-state structure of the same-chirality scattering theory. We argue that these constraints provide an efficient way to determine the mixed-chirality amplitudes.

The same-chirality amplitudes must contain the physical-strip poles associated with the bound states of the corresponding massive integrable theory. Consistency requires that these poles, together with their fusion processes, close under the $S$-matrix bootstrap. By contrast, the mixed-chirality amplitudes $S_{ab}^{\rm RL}$ should have no poles in the physical strip, since oppositely moving massless particles cannot form stable bound states. Imposing the bootstrap equations while excluding such poles leads to several distinct solutions for the left--right scattering amplitudes. For each solution, one can derive the corresponding thermodynamic Bethe ansatz and $Y$-system, determine its ultraviolet behavior, and identify the resulting RG flow and UV conformal field theory. This program has led to a classification of the UV-complete massless flows terminating at the $A_n$ coset CFTs \cite{Ahn1} and at the $E_8$ Ising theory \cite{AhnKim}.

The approach of \cite{AhnLeC} formulated the problem of ultraviolet completions of integrable irrelevant deformations in terms of massless left-right scattering amplitudes and their TBA equations. 
In the present work, we extend this framework to non-unitary minimal models and, more importantly, impose the full bound-state bootstrap on the mixed-chirality amplitudes \(S^{RL}_{ab}\). 
Together with crossing-unitarity and analyticity, these constraints allow us to construct the admissible amplitudes systematically from the known massive scattering theory. 
The resulting TBA equations identify an infinite family of RG flows from higher-fusion-level minimal models to non-unitary Virasoro minimal models, providing explicit realizations of ultraviolet completions beyond those previously studied.

 Although unitarity is essential for the standard probabilistic interpretation of quantum field theory, non-unitary CFTs play an important role in statistical mechanics, condensed-matter physics, and mathematical physics. They retain many of the computational techniques and algebraic structures of unitary CFTs, while exhibiting qualitatively different phenomena. In particular, although there is no general analogue of Zamolodchikov's $c$-theorem for non-unitary theories, the effective central charge $ c_{\mathrm{eff}}=c-24h_{\min} $ often plays an analogous role in rational models and decreases along many known non-unitary RG flows.

The simplest example of a non-unitary CFT is the Yang--Lee theory, which describes the edge singularity of the Ising model at imaginary magnetic field. More generally, non-unitary CFTs arise in systems with quenched disorder, non-Hermitian dynamics, quantum Hall criticality, and logarithmic conformal symmetry. They therefore provide both a rich mathematical framework and an effective description of a broad range of critical phenomena.

RG flows between non-unitary CFTs have previously been studied using conformal perturbation theory \cite{ZamRG} and effective Ginzburg--Landau descriptions \cite{Klebanov}. A well-known family is
\beq
\mathcal{M}(p,q)\longrightarrow \mathcal{M}(2p-q,p),
\qquad q>p,
\label{zamRGflow}
\eeq
first discussed in Refs.~\cite{AhnRG,Lassig}. Within this family, the minimal models $\mathcal{M}(2,2n+3)$, with $n$ a positive integer, are terminal: the same construction does not lead to another admissible minimal model. They therefore provide natural infrared fixed points for the inverse problem of constructing massless scattering theories and identifying their ultraviolet completions.

We show that a specific family of coset CFTs—the higher-fusion-level minimal models $\mathcal{M}(3,2n+3;n)$, perturbed by an appropriate relevant operator—flows to the Virasoro minimal models $\mathcal{M}(2,2n+3)$. The proposed flows are supported by complete solutions of the massless $S$-matrix bootstrap and by analytical and numerical comparisons between the resulting TBA systems and conformal perturbation theory around the ultraviolet fixed points.

Recent developments have revealed the importance of non-invertible symmetries in quantum field theory, particularly in two dimensions. Unlike ordinary global symmetries, they are generated by topological defect lines that need not possess fusion inverses and instead obey non-group-like fusion rules. Such structures occur in both unitary and non-unitary theories, but non-unitary CFTs provide a particularly broad setting in which non-unitary fusion categories and unconventional defect spectra may arise. Non-invertible symmetries have also proved useful in constraining RG dynamics and, in particular, in identifying possible flows between non-unitary conformal fixed points \cite{Nakayama}.

For the flows considered here, we identify a family of non-invertible Verlinde defect lines that is preserved by the perturbation. Equivalently, these defects act trivially on the perturbing operator after normalization and therefore remain topological along the RG trajectory. They can consequently be followed away from criticality and matched between the ultraviolet and infrared fixed points, providing an additional non-perturbative characterization of the proposed flows.

This paper is organized as follows. In Sect.~2, we determine the admissible left--right scattering amplitudes by solving the massless $S$-matrix bootstrap equations. In Sect.~3, we construct the corresponding TBA systems, analyze their ultraviolet behavior, and identify the associated UV-complete theories. In Sect.~4, we study the non-invertible symmetries preserved along the flows. We conclude in Sect.~5. Appendix~A collects the necessary background on non-unitary coset CFTs.

\section{Integrable deformations of the non-unitary minimal models
$\mathcal{M}({2,2n+3})$}

We begin by reviewing the exact massive scattering theory obtained by perturbing the minimal model $\mathcal{M}(2,2n+3)$ by the relevant operator $\phi_{1,3}$. Taking an appropriate massless scaling limit yields an integrable scattering description of the infrared CFT in which the same-chirality amplitudes $S^{\rm RR}$ and $S^{\rm LL}$ are inherited from the massive theory. Integrable irrelevant deformations of the IR fixed point couple the left- and right-moving sectors and generate nontrivial mixed-chirality amplitudes $S^{\rm RL}$ and $S^{\rm LR}$. These amplitudes must satisfy crossing-unitarity, and the bootstrap equations inherited from the bound-state structure of the massive theory. Within the class of solutions approaching finite constants at large rapidity, we classify all admissible mixed-chirality scattering amplitudes.

\subsection{Massive $S$-matrices}

We are interested in massless RG flows whose infrared endpoints are the non-unitary Virasoro minimal models $\mathcal{M}({2,2n+3})$. Recall that the minimal model $\mathcal{M}({p,q})$, with coprime integers $2\leq p<q$, has central charge and primary conformal weights
\beq
c=1-\frac{6(q-p)^2}{pq},
\qquad
h_{r,s}
=
\frac{(rq-sp)^2-(q-p)^2}{4pq},
\qquad
1\leq r\leq p-1,
\quad
1\leq s\leq q-1.
\eeq
The series $\mathcal{M}(2,2n+3)$ is particularly simple: after imposing the Kac-table identification
$ (r,s)\sim(p-r,q-s),$
there is a single independent row containing $n+1$ inequivalent primary fields, including the identity. The first member of the series, $\mathcal{M}(2,5)$, is the Yang--Lee edge-singularity CFT and contains only two primary fields.

The perturbation
\beq
\mathcal{M}(2,2n+3)
+
\lambda\int d^2x \, \phi_{1,3}(x),
\qquad
h_{1,3}
=
-\frac{2n-1}{2n+3},
\label{massivepert}
\eeq
preserves integrability. 
The resulting massive theory may be obtained by a quantum-group restriction of the sine-Gordon model at appropriate rational coupling, following the construction of Smirnov \cite{Smirnov} and Reshetikhin and Smirnov \cite{ResSmi}. 
Under this restriction, the soliton and antisoliton sectors are projected out, while \(n\) neutral breather particles \(B_a\), \(a=1,\ldots,n\), remain. 
Their masses are
\beq
m_a=M\sin\frac{a\pi}{h},
\qquad
h=2n+1,
\qquad
a=1,\ldots,n,
\label{masses}
\eeq
where $M$ sets the overall mass scale. The integer $h=2n+1$ is the Coxeter parameter associated with the twisted affine algebra $A_{2n}^{(2)}$, whose fusion structure underlies the scattering theory.
Their scattering amplitudes constitute the minimal \(A_{2n}^{(2)}\) S-matrix, whose fusion structure will play a central role in the massless bootstrap below.

For every pair $B_a,B_b$, the direct-channel pole at
\beq
u_{ab}^{c}=\frac{(a+b)\pi}{h}
\label{directfusionangle}
\eeq
corresponds to the bound state
\beq
c=
\begin{cases}
a+b, & a+b\leq n, \\
h-a-b, & a+b\geq n+1.
\end{cases}
\label{fusionrule}
\eeq
In the second case, the particle labels obey $a+b+c=h$. The fusing angle satisfies the on-shell relation
\beq
m_c^2
=
m_a^2+m_b^2+2m_am_b\cos u_{ab}^{c}.
\label{massshell}
\eeq
We also introduce the complementary angles
\(
\overline{u}_{ab}^{c}\equiv\pi-u_{ab}^{c}.
\)
The remaining angles in a fusion triangle are fixed by the analogous mass-shell relations and by
\beq
\overline{u}_{ab}^{c}
+
\overline{u}_{bc}^{a}
+
\overline{u}_{ca}^{b}
=
\pi.
\eeq

Let $\mathbb{A}_a(\theta)$ denote the Faddeev--Zamolodchikov operator associated with the particle $B_a$. Since the scattering is diagonal, these operators satisfy
\bea
\mathbb{A}_a(\theta_1)\mathbb{A}_b(\theta_2)
=
S_{ab}(\theta_1-\theta_2)\,
\mathbb{A}_b(\theta_2)\mathbb{A}_a(\theta_1).
\label{defS}
\eea
If $B_c$ is a bound state of $B_a$ and $B_b$, the corresponding fusion relation takes the schematic form
\beq
\mathbb{A}_a
\left(
\theta+i\overline{u}_{ac}^{b}
\right)
\mathbb{A}_b
\left(
\theta-i\overline{u}_{bc}^{a}
\right)
\ \sim
\Gamma_{ab}^{c}\,
\mathbb{A}_c(\theta),
\label{fusion}
\eeq
where $\Gamma_{ab}^{c}$ is the on-shell three-particle coupling. Combining \eqref{defS} and \eqref{fusion} gives the bootstrap equations
\bea
S_{dc}(\theta)
=
S_{da}
\left(
\theta-i\overline{u}_{ca}^{b}
\right)
S_{db}
\left(
\theta+i\overline{u}_{bc}^{a}
\right),
\qquad
a,b,c,d=1,\ldots,n.
\label{Sfusion}
\eea

The minimal solution of the bootstrap equations with the required physical-strip pole structure is
\beq
S_{ab}(\theta)
=
[a+b]_\theta\,
[|a-b|]_\theta
\prod_{k=1}^{\min(a,b)-1}
[|a-b|+2k]_\theta^{\,2},
\qquad
a,b=1,\ldots,n,
\label{breatherS}
\eeq
where
\beq
(x)_\theta
=
\frac{
\sinh\frac{1}{2}
\left(
\theta+\frac{i\pi x}{h}
\right)
}{
\sinh\frac{1}{2}
\left(
\theta-\frac{i\pi x}{h}
\right)
},
\qquad
[x]_\theta
=
-(x)_\theta(h-x)_\theta,
\qquad
[0]_\theta=1.
\label{blocks}
\eeq
These amplitudes constitute the minimal $S$-matrix associated with the twisted affine algebra $A_{2n}^{(2)}$. They may be obtained by folding the $A_{2n}^{(1)}$ scattering theory \cite{KlaMel}:
\beq
S_{ab}^{A_{2n}^{(2)}}(\theta)
=
S_{ab}^{A_{2n}^{(1)}}(\theta)\,
S_{ab}^{A_{2n}^{(1)}}(i\pi-\theta),
\qquad
a,b=1,\ldots,n.
\label{A2n2}
\eeq
Here $S_{ab}^{A_{2n}^{(1)}}$ is obtained from \eqref{breatherS} by replacing every block $[x]_\theta$ by $(x)_\theta$. Equivalently, the folding operation implements the blockwise replacement
\beq
(x)_\theta
\longmapsto
[x]_\theta
=
-(x)_\theta(h-x)_\theta.
\eeq
This relation will allow us to obtain the solutions of the $A_{2n}^{(2)}$ bootstrap equations from the corresponding complete set of solutions for the $A_{2n}^{(1)}$ theory.

\subsection{Massless $S$-matrices}

The massive integrable theories admit further irrelevant deformations that preserve integrability. Given the higher-spin conserved currents, one can construct scalar composite operators $[T\overline{T}]_s$, with $[T\overline{T}]_1=T\overline{T}$, and consider the action
\beq
\mathcal{S}_{\boldsymbol{\alpha}}
=
\mathcal{S}_{{\rm CFT}_{\rm IR}}
+
\lambda\int d^2x\,\Phi_{\rm rel}
+
\sum_{s\in\mathcal I}\alpha_s
\int d^2x\,[T\overline{T}]_s .
\label{pertcftTT}
\eeq
Here $\mathcal I$ denotes the set of spins of the available local integrals of motion. The deformations generated by these operators modify the factorized scattering amplitudes only through scalar CDD factors \cite{SmiZam}.

We now take the standard massless scaling limit of the massive theory. We shift the rapidity according to
\beq
\theta=\widehat{\theta}\pm\Lambda,
\qquad
M\to0,
\qquad
\Lambda\to\infty,
\qquad
\frac{M}{2}e^\Lambda\equiv\mu
\quad\text{fixed}.
\label{masslesslimit}
\eeq
Using $m_a=M\sin(a\pi/h)$, the massive dispersion relation
\beq
(E_a,P_a)=(m_a\cosh\theta,m_a\sinh\theta)
\eeq
then becomes
\beq
\begin{aligned}
(E_a^{\rm R},P_a^{\rm R})
&=
\left(\mu_a e^{\widehat{\theta}},
\mu_a e^{\widehat{\theta}}\right),
\
(E_a^{\rm L},P_a^{\rm L})
&=
\left(\mu_a e^{-\widehat{\theta}},
-\mu_a e^{-\widehat{\theta}}\right),
\end{aligned}
\qquad
\mu_a=\mu\sin\frac{a\pi}{h}.
\label{masslessdispersion}
\eeq
The upper and lower signs in \eqref{masslesslimit} therefore produce right-moving and left-moving particles, respectively. We henceforth drop the hat from the massless rapidity.

The CDD deformation of a massive two-particle amplitude has the schematic form
\beq
S^{(\boldsymbol{\alpha})}_{ab}(\theta)
=
S_{ab}(\theta)
\exp\left[
i\sum_{s\in\mathcal I}
\alpha_s q_s^{(a)}q_s^{(b)}
\sinh(s\theta)
\right],
\label{massiveCDD}
\eeq
where $q_s^{(a)}$ is the one-particle eigenvalue of the conserved charge of spin $s$, and scales as $m_a^s$.

For two particles of the same chirality, the rapidity difference remains finite:
\beq
(\theta_1\pm\Lambda)-(\theta_2\pm\Lambda)
=
\theta_1-\theta_2.
\eeq
The CDD contribution in \eqref{massiveCDD} consequently vanishes as $M^{2s}$, while the undeformed massive amplitude remains finite. Hence
\beq
S_{ab}^{\rm RR}(\theta)
= S_{ab}^{\rm LL}(\theta)=
S_{ab}(\theta).
\label{samechirality}
\eeq
The same-chirality amplitudes therefore retain the complete bound-state and bootstrap structure of the massive theory.

For particles of opposite chirality, the massive rapidity difference behaves instead as
\(
(\theta_1+\Lambda)-(\theta_2-\Lambda)
=
\theta_1-\theta_2+2\Lambda.
\)
The undeformed massive amplitude approaches a constant phase, which can be absorbed into the definition of the asymptotic states. On the other hand,
\(
q_s^{(a)}q_s^{(b)}
\sinh\left[s(\theta+2\Lambda)\right]
\)
has a finite limit because \(q_s^{(a)}q_s^{(b)}\sim M^{2s}\) while
\(e^{2s\Lambda}\sim M^{-2s}\). The mixed-chirality amplitudes therefore take the form
\beq
S^{\rm RL}_{ab}(\theta)
= S^{\rm LR}_{ab}(-\theta)=
\exp\left(
i\sum_{s\in\mathcal I}
g_s^{ab}e^{s\theta}
\right),
\label{SabTT}
\eeq
where the coefficients \(g_s^{ab}\) are determined by the couplings \(\alpha_s\) and by the charge eigenvalues. Before imposing analyticity, crossing-unitarity, and the bootstrap equations, these parameters leave considerable freedom in the left--right scattering amplitudes.

The Faddeev--Zamolodchikov operators also split into right- and left-moving copies,
\beq
\mathbb{A}^{\rm R/L}_a(\theta)
=
\lim_{\Lambda\to\infty}
\mathbb{A}_a(\theta\pm\Lambda).
\eeq
Since the rapidity shifts cancel within each chiral sector, the fusion relations are inherited directly from the massive theory:
\beq
\mathbb{A}^{\rm R/L}_a
\left(\theta+i\overline{u}^{b}_{ac}\right)
\mathbb{A}^{\rm R/L}_b
\left(\theta-i\overline{u}^{a}_{bc}\right)
\ \sim
\Gamma_{ab}^{c}\,
\mathbb{A}^{\rm R/L}_c(\theta).
\label{fusionmassless}
\eeq

The mixed-chirality scattering amplitudes are defined by the exchange relations
\bea
\mathbb{A}^{\rm R}_a(\theta_1)
\mathbb{A}^{\rm L}_b(\theta_2)
&=&
S^{\rm RL}_{ab}(\theta_1-\theta_2)\,
\mathbb{A}^{\rm L}_b(\theta_2)
\mathbb{A}^{\rm R}_a(\theta_1)\,
\nonumber\\
\mathbb{A}^{\rm L}_a(\theta_1)
\mathbb{A}^{\rm R}_b(\theta_2)
&=&
S^{\rm LR}_{ab}(\theta_1-\theta_2)\,
\mathbb{A}^{\rm R}_b(\theta_2)
\mathbb{A}^{\rm L}_a(\theta_1).
\label{defSRL}
\eea
Combining these relations with \eqref{fusionmassless} gives
\bea
S^{\rm RL}_{dc}(\theta)
=
S^{\rm RL}_{da}
\left(\theta-i\overline{u}^{b}_{ca}\right)
S^{\rm RL}_{db}
\left(\theta+i\overline{u}^{a}_{bc}\right),
\qquad
a,b,c,d=1,\ldots,n,
\label{Sfusionmassless}
\eea
and an analogous equation for \(S^{\rm LR}_{ab}\). Thus, the mixed-chirality amplitudes obey the same algebraic bootstrap equations as the massive amplitudes.

Because all particles in the \(A_{2n}^{(2)}\) scattering theory are self-conjugate, crossing and algebraic unitarity combine into
\beq
S^{\rm RL}_{ab}(\theta)
S^{\rm RL}_{ab}(\theta+i\pi)
=
1.
\label{crossingunitarityRL}
\eeq
An additional essential requirement is that \(S^{\rm RL}_{ab}\) have no poles in the physical strip
$ 0<\operatorname{Im}\theta<\pi,$
since a right-moving and a left-moving massless particle cannot form a stable bound state.

The bootstrap equations admit several solutions that are closely related to those of the massive theory. The simplest is the diagonal solution
\beq
S^{\rm diag}_{ab}(\theta)
=
S_{ab}^{-1}(\theta),
\label{diagonalS}
\eeq
which satisfies the same bootstrap and crossing-unitarity equations. Taking the inverse converts the physical-strip bound-state poles of the massive amplitudes into zeros and therefore produces an admissible pole-free mixed-chirality solution.

A systematic classification follows from the folding relation
\eqref{A2n2}. The elementary solutions of the \(A_{2n}^{(2)}\)
bootstrap equations can be obtained by folding the corresponding
solutions of the \(A_{2n}^{(1)}\) theory classified in
Ref.~\cite{Ahn1}. After folding, the independent elementary solutions
may be labelled by
\beq
S_{11}^{(k)}(\theta)=[-k]_\theta,
\qquad k=1,\ldots,n,
\eeq
with all higher-particle amplitudes fixed by the bootstrap. The
\(k=1\) solution is the minimal amplitude discussed below, while
\(k=2\) gives the diagonal solution \eqref{diagonalS}; products of
the elementary generators give further solutions, including the
saturated amplitude \((S^{\rm min})^2\).  

These amplitudes solve the algebraic bootstrap problem, but this does
not by itself guarantee a conformal ultraviolet completion. As we
discuss in Subsect.~\ref{subsec:othersolutions}, the requirement that
the corresponding TBA possess a positive constant (Y)-system
solution imposes an additional and considerably stronger restriction.

For \(A_{2n}^{(1)}\), the minimal amplitudes are
\beq
S_{ab}^{\rm min}[A_{2n}^{(1)}](\theta)
=
\left [ (a+b-1)_\theta
(a+b-3)_\theta
\cdots
(|a-b|+3)_\theta
(|a-b|+1)_\theta \right ]^{-1} .
\eeq
Applying the folding prescription block by block gives the mixed-chirality amplitudes
\beq
S_{ab}^{\rm min}(\theta)
=
\left ( [a+b-1]_\theta
[a+b-3]_\theta
\cdots
[|a-b|+3]_\theta
[|a-b|+1]_\theta \right )^{-1} ,
\qquad
a,b=1,\ldots,n.
\label{minimalS}
\eeq
These amplitudes satisfy the massless bootstrap equations
\eqref{Sfusionmassless}, crossing-unitarity
\eqref{crossingunitarityRL}, and contain no poles in the physical strip. They  correspond to a resummation of an infinite hierarchy of generalized $T \bar T$-type deformations.

\section{Thermodynamic Bethe ansatz}

We now analyze the TBA equations associated
with the minimal left--right scattering amplitudes \eqref{minimalS}.
We first rewrite the equations in a universal Dynkin-diagram form and
derive the corresponding \(Y\)-system. We then study the small-volume,
or ultraviolet, limit. The plateau solutions determine the UV effective
central charge, while the periodicity of the \(Y\)-system determines the
scaling dimension of the perturbing operator. Finally, we extract the
leading finite-size corrections analytically and compare them with
conformal perturbation theory around the proposed UV fixed point. These
tests allow us to identify both the UV CFT and the relevant perturbation
that generates the flow.

\subsection{TBA for the minimal \(S\)-matrix}

Since all scattering amplitudes are diagonal, the ground-state TBA
equations follow directly from the complete set of amplitudes
\(S^{\rm RR}_{ab}\), \(S^{\rm LL}_{ab}\), \(S^{\rm RL}_{ab}\), and
\(S^{\rm LR}_{ab}\). On a circle of circumference \(\mathcal R\), the
right- and left-moving pseudoenergies satisfy
\beq
\epsilon_a^{\rm R/L}(\theta)
=
d_a^{\rm R/L}(\theta)
-\sum_{b=1}^{n}
 \phi_{ab}\star L_b^{\rm R/L}(\theta)
-\sum_{b=1}^{n}
 \psi_{ab}\star L_b^{\rm L/R}(\theta),
\label{rawTBA}
\eeq
where
\beq
L_a^{\rm R/L}(\theta)
=
\log\left(1+e^{-\epsilon_a^{\rm R/L}(\theta)}\right).
\label{defL}
\eeq
The driving terms are
\beq
d_a^{\rm R}(\theta)
=
\frac{r\,m_a}{2m_1}e^\theta,
\qquad
d_a^{\rm L}(\theta)
=
\frac{r\,m_a}{2m_1}e^{-\theta},
\qquad
r=m_1\mathcal R .
\label{drivingterms}
\eeq
The convolution is defined by
\beq
(f\star g)(\theta)
=
\int_{-\infty}^{\infty}\frac{d\theta'}{2\pi}\,
f(\theta-\theta')g(\theta'),
\eeq
and the kernels are logarithmic derivatives of the scattering
amplitudes,
\beq
\phi_{ab}(\theta)
=
-i\frac{d}{d\theta}
\log S_{ab}^{\rm RR}(\theta)
=
-i\frac{d}{d\theta}
\log S_{ab}^{\rm LL}(\theta)\quad ,\quad
\psi_{ab}(\theta)
=
-i\frac{d}{d\theta}
\log S_{ab}^{\rm RL}(\theta).
\label{TBAkernels}
\eeq
Parity invariance implies
\(
\epsilon_a^{\rm L}(\theta)
=
\epsilon_a^{\rm R}(-\theta),
\)
so it is sufficient in practice to solve one of the two sets of
equations.

The ground-state energy can be obtained from the pseudo-energies as 
\beq
E_0({\cal R})=-\sum_{a=1}^n \frac{m_a}{2}\int_{-\infty}^{\infty} \frac{d\theta}{2\pi} \left[L_a^R(\theta)e^\theta+L_a^L(\theta)e^{-\theta} \right]
\label{E0}
\eeq

For the analytic treatment, it is convenient to introduce Fourier
transforms with conventions such that convolution becomes
multiplication. Defining
\(
m_{ab}=\min(a,b),
\) and \(
M_{ab}=\max(a,b),
\)
the Fourier-transformed kernels take the symmetric form
\bea
\widetilde{\phi}_{ab}(w)
&=&
\delta_{ab}
-
\frac{
2\cosh\frac{w}{h}\,
\sinh\frac{m_{ab}w}{h}\,
\cosh\left[\left(\frac{h}{2}-M_{ab}\right)\frac{w}{h}\right]
}{
\cosh\frac{w}{2}\,
\sinh\frac{w}{h}
},
\label{phifourier}
\\
\widetilde{\psi}_{ab}(w)
&=&
\frac{
\sinh\frac{m_{ab}w}{h}\,
\cosh\left[\left(\frac{h}{2}-M_{ab}\right)\frac{w}{h}\right]
}{
\cosh\frac{w}{2}\,
\sinh\frac{w}{h}
}.
\label{psifourier}
\eea
These matrices obey
\beq
\left[
\mathbf 1-\widetilde{\boldsymbol\phi}(w)
\right]^{-1}
=
\mathbf 1-
\widetilde{\varphi}_h(w)\,\mathcal I\quad ,\quad
\left[
\mathbf 1-\widetilde{\boldsymbol\phi}(w)
\right]^{-1}
\widetilde{\boldsymbol\psi}(w)
=
\widetilde{\varphi}_h(w)\,\mathbf 1,
\label{kernelidentity}
\eeq
where
\beq
\widetilde{\varphi}_h(w)
=
\frac{1}{2\cosh\frac{w}{h}},
\qquad
\varphi_h(\theta)
=
\frac{h}{2\cosh\frac{h\theta}{2}}.
\label{universalKernel}
\eeq
The matrix
\beq
\mathcal I_{ab}
=
\delta_{a,b-1}
+
\delta_{a,b+1}
+
\delta_{a,n}\delta_{b,n},
\qquad
a,b=1,\ldots,n,
\label{incidencematrix}
\eeq
is the incidence matrix of the \(n\)-node tadpole graph
\(
T_n=A_{2n}/\mathbb Z_2,
\)
which is naturally associated with the \(A_{2n}^{(2)}\) scattering
theory. In particular, the mass ratios form its Perron--Frobenius
eigenvector:
\(
\sum_{b=1}^{n}\mathcal I_{ab}m_b
=
2\cos\frac{\pi}{h}\,m_a.
\)

To transform \eqref{rawTBA}, we introduce
\beq
\overline L_a^{\rm R/L}
=
\log\left(1+e^{\epsilon_a^{\rm R/L}}\right)
=
\epsilon_a^{\rm R/L}+L_a^{\rm R/L}.
\label{defbarL}
\eeq
Using \eqref{kernelidentity}, the TBA
equations become
\bea
\epsilon_a^{\rm R/L}
=
d_a^{\rm R/L}
+
\sum_{b=1}^{n}
\mathcal I_{ab}\,
\varphi_h\star
\left(
\overline L_b^{\rm R/L}-d_b^{\rm R/L}
\right)
-
\varphi_h\star L_a^{\rm L/R} .
\label{universalTBA}
\eea
The subtraction of the driving terms inside the convolutions makes
\eqref{universalTBA} convergent and is essential when deriving the
universal form directly from the Fourier-space identities.

The system can be represented by two copies of the tadpole diagram
\(T_n\), corresponding to the right- and left-moving pseudoenergies.
Adjacent nodes within each copy are coupled through
\(\mathcal I_{ab}\), while every right-moving node is coupled to the
left-moving node carrying the same particle label. This Dynkin form
provides the natural starting point for deriving the \(Y\)-system and
analyzing the ultraviolet limit. Indeed, applying the inverse of the universal kernel to the simplified TBA equations leads to a closed set of functional relations. 
Introducing
\beq
Y_a^{\rm R/L}(\theta)=e^{\epsilon_a^{\rm R/L}(\theta)},
\eeq
and using the identity
\beq
(\phi_h\star f)\left(\theta+i\pi/h\right)
+
(\phi_h\star f)\left(\theta-i\pi/h\right)
=
f(\theta),
\eeq
we obtain the $Y$-system
\beq
Y_a^{\rm R/L}\left(\theta+i\pi/h\right)
Y_a^{\rm R/L}\left(\theta-i\pi/h\right)
=
\frac{
\prod_{b=1}^{n}
\left(1+Y_b^{\rm R/L}(\theta)\right)^{\mathcal{I}_{ab}}
}{
1+1/Y_a^{\rm L/R}(\theta)
},
\label{Ysystem}
\eeq
In deriving these relations, the driving terms cancel because the mass vector is an eigenvector of the incidence matrix. 
The resulting $Y$-system consists of two copies of the $A_{2n}^{(2)}$ tadpole diagram, coupled node by node through the denominators in \eqref{Ysystem}. In contrast to the original TBA equations, these functional relations contain no explicit dependence on the scale $r$. Their constant solutions determine the ultraviolet effective central charge, while their periodicity in imaginary rapidity determines the conformal dimension of the relevant operator generating the flow.

The $Y$-system can be regarded as a coupled second-order difference equation in the imaginary rapidity direction. Given the values of \(Y_a^{\rm L/R}\) at \(\theta\) and \(\theta+i\pi/h\), the functional relations recursively determine their values at
\(
\theta+\frac{i\pi j}{h}\) for \(j=2,3,\ldots\). By starting with some initial 
numerical values and investigating their recurrence behavior
we find that the system is periodic after
\(
N_n=4n+8
\)
such steps:
\beq
Y_a^{\rm L/R}(\theta)
=
Y_a^{\rm L/R}\left(
\theta+N_n i\pi/h
\right)
=
Y_a^{\rm L/R}\left(\theta+i\pi P_n\right),
\qquad
P_n=\frac{N_n}{h}
=\frac{4n+8}{2n+1}.
\label{periodicity}
\eeq
Using the relation between the period of a massless $Y$-system and the chiral dimension of the UV perturbing relevant operator,
\(
P_n=1/(1-h_{\rm rel}),
\)
we obtain
\beq
h_{\rm rel}
=1-\frac{1}{P_n}=
\frac{2n+7}{4n+8}.
\eeq
In the following, we complement this result by extracting the UV effective central charge from the small-volume expansion of the TBA equations.

\begin{figure}[t]
\centering
\begin{tikzpicture}[
node distance=1.2cm and 1.4cm,
ynode/.style={
circle,
draw,
minimum size=6.5mm,
inner sep=0pt
},
hlink/.style={draw, thick},
vlink/.style={draw, thick, dashed}
]

\node[ynode] (R1) {$1$};
\node[ynode, right=of R1] (R2) {$2$};
\node[right=of R2] (Rdot) {$\cdots$};
\node[ynode, right=1.6cm of Rdot] (Rn) {$n$};

\node[ynode, below=1.4cm of R1] (L1) {$1$};
\node[ynode, below=1.4cm of R2] (L2) {$2$};
\node[below=1.6cm of Rdot] (Ldot) {$\cdots$};
\node[ynode, below=1.4cm of Rn] (Ln) {$n$};

\draw[hlink] (R1)--(R2);
\draw[hlink] (R2)--(Rdot);
\draw[hlink] (Rdot)--(Rn);

\draw[hlink] (L1)--(L2);
\draw[hlink] (L2)--(Ldot);
\draw[hlink] (Ldot)--(Ln);

\path (Rn) edge[hlink, loop right, min distance=16mm] (Rn);
\path (Ln) edge[hlink, loop right, min distance=16mm] (Ln);

\draw[vlink] (R1)--(L1);
\draw[vlink] (R2)--(L2);
\draw[vlink] (Rn)--(Ln);

\node[left=0.6cm of R1] {\(\mathrm{R}\)};
\node[left=0.6cm of L1] {\(\mathrm{L}\)};

\end{tikzpicture}
\caption{Diagrammatic representation of the doubled tadpole \(Y\)-system.
Each row is a copy of the tadpole graph \(T_n=A_{2n}/\mathbb Z_2\), and
dashed vertical links couple nodes with the same label in the right- and
left-moving sectors.}
\label{fig:doubledYsystem}
\end{figure}
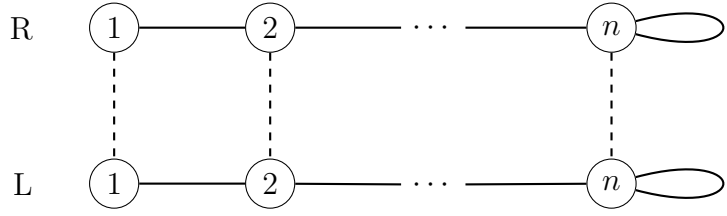

\subsection{Small-volume expansion}

In the UV limit, $r\to 0$, the ground-state energy obtained from the TBA \eqref{E0} takes the form
\beq
\label{uvexpansion}
E_0(\mathcal{R})
=
\frac{2\pi}{\mathcal{R}}
\left(
-\frac{c_{\mathrm{eff}}}{12}
+
\sum_{j=1}^{\infty}
c_j
\left(\frac{r}{2\pi}\right)^{j(2-2h_{\mathrm{rel}})}
\right)-
E_{\mathrm{bulk}},
\eeq
where $c_{\mathrm{eff}}$ is the effective central charge of the UV theory and $h_{\mathrm{rel}}$ is the chiral dimension of the relevant perturbing operator. The term $E_{\mathrm{bulk}}$ is related to the bulk-energy constant and accounts for the difference between the TBA and perturbed-CFT normalization schemes. Its negative sign reflects the fact that the TBA ground-state energy is normalized to vanish at large volume, whereas the perturbed-CFT ground-state energy admits a regular expansion in $\mathcal{R}^{2-2h_{\mathrm{rel}}}$. The last term is commonly referred to as the anti-bulk contribution.

Our aim is to expand the TBA equations and extract these quantities. This will help us identify the UV theory from which the flow to the infrared $\mathcal{M}(2,2n+3)$ minimal-model CFT originates.

To simplify the notation, we introduce a unified notation for the left- and right-moving particles. Capital letters denote multi-indices
\(
I=1,\ldots,n,n+1,\ldots,2n,
\)
such that the first $n$ values correspond to right-moving particles and the last $n$ to left-moving particles. In particular,
\[
m_I=
\begin{cases}
m_I & I\leq n\\
m_{I-n} & I>n
\end{cases} \quad ;\quad \epsilon_I=
\begin{cases}
\epsilon_I^{\rm R} & I\leq n \\
\epsilon_{I-n}^{\rm L}  & I>n
\end{cases}.
\]
The same convention is used for $L_I$ and $d_I$. In this ordering, the full $S$-matrix has the block structure
\beq
S=
\begin{pmatrix}
S^{\rm RR} & S^{\rm RL}\\
S^{\rm LR} & S^{\rm LL}
\end{pmatrix},
\eeq
with the corresponding kernel $\varphi$ defined by its logarithmic derivative. The TBA equations then take the compact form
\beq
\epsilon_I=d_I-\varphi_{IJ}\star L_J.
\eeq
It is convenient to introduce the scaling function, which in the multi-index notation reads
\beq
c(r)
=-\frac{6\mathcal{R}}{\pi}E_0(\mathcal{R})=
\sum_{I=1}^{2n}
\infint
\frac{3\,d\theta}{\pi^2}\,
d_I L_I .
\eeq

\begin{figure}[h]
\centering
\includegraphics[width=0.5\linewidth]{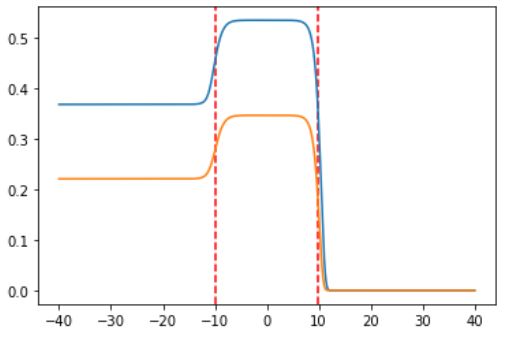}
\caption{Functional form of the right-moving $L$-functions for $n=2$. The kink positions $\theta=\pm\Lambda$ are indicated by red dashed lines. The left-moving functions are obtained by reflection about the $y$-axis.}
\label{fig:Lshape}
\end{figure}

At very small volume, the shape of the $L$-functions shown in Fig.~\ref{fig:Lshape} indicates that their nontrivial variation is localized in the kink region
\(
\theta\sim\Lambda:=-\log(r/2)
\)
and the antikink region $\theta\sim-\Lambda$. In the kink region, corresponding to large positive rapidities, the driving terms of the left-moving particles are exponentially suppressed and can be neglected. The kink TBA is therefore defined by
\beq
\epsilon_I^K=d_I^K-\varphi_{IJ}\star L_J^K,
\eeq
where
\(
d_I^K=
d_I \) for $ I\leq n$ and $0$ otherwise. 
The kink functions depend on the volume only through a shift of their arguments and can be expressed in terms of volume-independent functions as
\beq
\epsilon_I^K(\theta)
=
\epsilon_I^+\left(\theta+\log\frac{r}{2}\right),
\qquad
L_I^K(\theta)
=
L_I^+\left(\theta+\log\frac{r}{2}\right),
\eeq
where $\epsilon_I^+$ and $L_I^+$ satisfy the same equations as $\epsilon_I^K$ and $L_I^K$, but with $r=2$.

The antikink TBA is defined analogously and describes the large negative-rapidity behavior of the functions $\epsilon_I$ and $L_I$. In the antikink region, the driving terms are
$
d_I^A=0 $ for $ I\leq n $ and $ d_I^A=d_I$ otherwise. 
The antikink functions can be expressed in terms of their volume-independent counterparts as
\beq
\epsilon_I^A(\theta)
=
\epsilon_I^-\left(\theta-\log\frac{r}{2}\right),
\qquad
L_I^A(\theta)
=
L_I^-\left(\theta-\log\frac{r}{2}\right).
\eeq

\begin{figure}[h]
\centering
\includegraphics[width=0.5\linewidth]{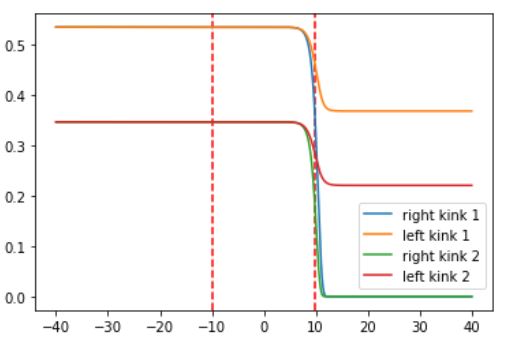}
\caption{Functional form of the kink functions for $n=2$. The corresponding antikink functions are obtained by reflection about the $y$-axis together with the interchange of right- and left-moving particles.}
\label{fig:Lkink}
\end{figure}

The full functions can be reconstructed from the kink and antikink solutions up to small corrections:
\beq
\epsilon_I
=
\epsilon_I^K+\epsilon_I^A-\epsilon_I^0+\delta\epsilon_I,
\qquad
L_I
=
L_I^K+L_I^A-L_I^0+\delta L_I,
\eeq
where $\epsilon_I^0$ and $L_I^0$ are the plateau values of the pseudoenergy and the $L$-function, respectively.

The kink solution $L_I^K$ contributes to the central charge, $L_I^A-L_I^0$ to the bulk energy, while the small corrections $\delta\epsilon_I$ and $\delta L_I$ generate the perturbative contributions of order $r^{j(2-2h)}$ to the ground-state energy. At leading order, they can be determined from a linear integral equation, as discussed in Subsect.~\ref{LOpert}.

\subsection{UV effective central charge}

The UV effective central charge can be expressed in terms of the kink functions as
\beq
c_{\mathrm{eff}}^{\mathrm{UV}}
=
\sum_{I=1}^{2n}
\int_{-\infty}^{\infty}
\frac{6\,d\theta}{\pi^2}\,
d_I^+(\theta)L_I^+(\theta).
\label{ceffkink}
\eeq
Here we used the equality of the kink and antikink contributions, shifted the rapidity to eliminate the explicit volume dependence, and extended the sum to all multi-indices by setting \(d_I^+=0\) for \(I>n\).

The standard dilogarithm argument proceeds by using
\(
\partial_\theta d_I^+=d_I^+
\)
together with the derivative of the kink TBA equations. After an integration by parts and using the symmetry of the kernel, one finds
\beq
c_{\mathrm{eff}}^{\mathrm{UV}}
=
\sum_I
\int_{-\infty}^{\infty}
\frac{6\,d\theta}{\pi^2}\,
\partial_\theta\epsilon_I^+
\left[
L_I^+
-
\frac{d_I^+-\epsilon_I^+}{1+e^{\epsilon_I^+}}
\right].
\eeq
The term proportional to \(d_I^+\) reproduces the original integral, and hence
\beq
c_{\mathrm{eff}}^{\mathrm{UV}}
=
\sum_I
\int_{-\infty}^{\infty}
\frac{3 d\theta}{\pi^2}\,
\partial_\theta\epsilon_I^+
\left[
\log\left(1+e^{-\epsilon_I^+}\right)
+
\frac{\epsilon_I^+}{1+e^{\epsilon_I^+}}
\right]
=
-\frac{6}{\pi^2}
\sum_I
\int_{-\infty}^{\infty}
d\theta\,
\partial_\theta
\mathcal L\left(
\frac{1}{1+e^{\epsilon_I^+}}
\right),
\label{ceffdilog}
\eeq
where
\beq
\mathcal L(x)
=
\operatorname{Li}_2(x)
+
\frac{1}{2}\log x\log(1-x)
\eeq
is the Rogers dilogarithm.

For the right-moving particles, \(\epsilon_i^+(\theta)\to\infty\) as
\(\theta\to\infty\), so the upper boundary gives no contribution. Their contribution is therefore
\beq
\frac{6}{\pi^2}
\sum_{i=1}^n
\mathcal L\left(
\frac{1}{1+Y_i^0}
\right),
\qquad
Y_I^0=e^{\epsilon_I^0},
\label{yplateau}
\eeq
where the common plateau values satisfy
\beq
Y_I^0
=
\prod_{J=1}^{2n}
\left(1+\frac{1}{Y_J^0}\right)^{-\widehat\varphi_{IJ}},
\qquad
\widehat\varphi_{IJ}
=
\int_{-\infty}^{\infty}
\frac{d\theta}{2\pi}\,
\varphi_{IJ}(\theta).
\label{masslessYsystem}
\eeq
Parity implies
\(
\epsilon_i^{\rm L}(\theta)
=\epsilon_i^{\rm R}(-\theta),
\)
so the right- and left-moving plateau values coincide.

At the opposite end of the left-moving kink, the right-moving particles decouple and the pseudoenergies approach the plateau values \(\widetilde Y_i^0\) of the massive \(A_{2n}^{(2)}\) scattering theory:
\beq
\widetilde Y_i^0
=
\prod_{j=1}^n
\left(1+1/{\widetilde Y_j^0}\right)^{-\widehat\phi_{ij}}.
\label{massiveYsystem}
\eeq
The corresponding local constant $Y$-system is that of the tadpole graph
\(T_n=A_{2n}/\mathbb Z_2\):
\bea
(\widetilde Y_a^0)^2
&=&
(1+\widetilde Y_{a-1}^0)
(1+\widetilde Y_{a+1}^0),
\qquad
a=1,\ldots,n-1,
\nonumber\\
(\widetilde Y_n^0)^2
&=&
(1+\widetilde Y_{n-1}^0)
(1+\widetilde Y_n^0),
\qquad
\widetilde Y_0^0=0.
\eea
Its positive solution is
\beq
\widetilde Y_a^0
=
\frac{
\sin(a\vartheta_{\mathrm{IR}})
\sin((a+2)\vartheta_{\rm IR})
}{
\sin^2\vartheta_{\rm IR}
}\quad ,
\qquad
\vartheta_{\rm IR}
=
\frac{\pi}{2n+3},
\qquad
a=1,\ldots,n.
\label{IRplateaus}
\eeq
These values reproduce the effective central charge of the infrared minimal model:
\beq
\frac{6}{\pi^2}
\sum_{i=1}^n
\mathcal L\left(
\frac{1}{1+\widetilde Y_i^0}
\right)
=1-\frac{6}{2(2n+3)}
=
c_{\mathrm{eff}}^{\mathrm{IR}}.
\label{ceffIR}
\eeq

For the doubled tadpole system, the constant equations are
\bea
\left(Y_a^{\rm L/R}\right)^2
&=&
\frac{
(1+Y_{a-1}^{\rm L/R})
(1+Y_{a+1}^{\rm L/R})
}{
1+1/Y_a^{\rm R/L}
},
\qquad
a=1,\ldots,n-1,
\nonumber\\
\left(Y_n^{\rm L/R}\right)^2
&=&
\frac{
(1+Y_{n-1}^{\rm L/R})
(1+Y_n^{\rm L/R})
}{
1+1/Y_n^{\rm R/L}
},
\qquad
Y_0^{\rm L/R}=0.
\label{doubledconstantY}
\eea
The positive solution is symmetric,
\(
Y_a^{\rm L}=Y_a^{\rm R}\equiv Y_a^0,
\)
and is given by
\bea
1+Y_a^0
&=&
\frac{
\sin((a+1)\vartheta_{\rm UV})
\sin((a+2)\vartheta_{\rm UV})
}{
\sin\vartheta_{\rm UV}\,
\sin(2\vartheta_{\rm UV})
}\quad ,
\nonumber\\
Y_a^0
&=&
\frac{
\sin(a\vartheta_{\rm UV})
\sin((a+3)\vartheta_{\rm UV})
}{
\sin\vartheta_{\rm UV}\,
\sin(2\vartheta_{\rm UV})
}\quad,
\qquad
\vartheta_{\rm UV}
=
\frac{\pi}{2n+4}.
\label{UVplateaus}
\eea

Combining the two chiral contributions at the common plateau and subtracting the massive plateau contribution from the upper boundary of the left-moving kink, we obtain
\beq
c_{\mathrm{eff}}^{\mathrm{UV}}
=
\frac{12}{\pi^2}
\sum_{i=1}^n
\mathcal L\left(
\frac{1}{1+Y_i^0}
\right)
-
c_{\mathrm{eff}}^{\mathrm{IR}}
=
\frac{3n}{n+2}-\frac{6n}{3(2n+3)}
.
\label{ceffUV}
\eeq

\subsection{Bulk-energy coefficient}

For \(n=1\), the perturbing dimension \(h_{\rm rel}=3/4\) implies
\(
2-2h_{\rm rel}=\frac{1}{2},
\)
and the fourth perturbative order resonates with the bulk contribution. The latter therefore takes the modified form
\beq
E_{\rm bulk}
=
\begin{cases}
\epsilon_{\rm bulk}^{(0)}\,r
+
\epsilon_{\rm bulk}^{(1)}\,r\log r
& n=1, \\
\epsilon_{\rm bulk}\,r
& n\geq2.
\end{cases}
\label{bulkform}
\eeq
The \(n=1\) case is exceptional because the pseudoenergy approaches its plateau value as \(e^{-|\theta|}\), producing the logarithmic term $\epsilon_{\rm bulk}^{(1)}=\frac{3}{4\pi^2} m_1 $ calculated in Ref.~\cite{RavaniniStanishkovTateo} along with $\epsilon_{\rm bulk}^{(0)}=\frac{\sqrt{3}}{4\pi} m_1 $. 

For \(n\geq2\), the bulk-energy coefficient can be extracted from the absence of \(e^{\pm\theta}\) terms in the expansion around the plateau. The bulk contribution is given by
\beq
\epsilon_{\rm bulk}\,r
=
\sum_{I=n+1}^{2n}m_I
\int_{-\infty}^{\infty}
\frac{d\theta}{2\pi}\,
e^{-\theta}
\left(L_I^K-L_I^0\right)
=
2\sum_{I=n+1}^{2n}m_I w_I,
\label{bulkIntegral}
\eeq
where
\beq
w_I
:=
\int_{-\infty}^{\infty}
\frac{d\theta}{2\pi}\,
e^{-\theta}\partial_\theta L_I^K(\theta).
\label{defw}
\eeq
In the second equality of \eqref{bulkIntegral}, we integrated by parts.

At large rapidity, the kernels behave as
\beq
\phi_{IJ}(\theta)
=
A_{IJ}e^{-|\theta|}
+
O\left(e^{-2|\theta|}\right).
\label{kernelAsymptotics}
\eeq
The nonvanishing coefficients relevant for the calculation are
\beq
A_{ab}
=
-8\cos\frac{\pi}{h}\,
\frac{
\sin\frac{a\pi}{h}\sin\frac{b\pi}{h}
}{
\sin\frac{\pi}{h}
}\quad ;\quad A_{a,b+n}
=
4\,
\frac{
\sin\frac{a\pi}{h}\sin\frac{b\pi}{h}
}{
\sin\frac{\pi}{h}
}
\label{Acoefficients}
\eeq
where $
a,b=1,\ldots n$. 
Consider first the \(\theta\to-\infty\) asymptotics of the right-moving kink equations. For \(i=1,\ldots,n\),
\beq
\partial_\theta\epsilon_i^K(\theta)
=
\left[
\frac{m_i\mathcal R}{2}
-
\sum_{J=1}^{2n}A_{iJ}w_J
\right]e^\theta
+\cdots .
\label{rightKinkAsymptotics}
\eeq
Since the plateau expansion contains no \(e^\theta\) term for \(n\geq2\), its coefficient must vanish:
\beq
\sum_{J=1}^{2n}A_{iJ}w_J
=
\frac{m_i\mathcal R}{2}.
\label{rightConstraint}
\eeq
The left-moving kink equations contain no driving term in this limit and similarly give
\beq
\sum_{J=1}^{2n}A_{i+n,J}w_J
=
0\quad,
\qquad
i=1,\ldots,n.
\label{leftConstraint}
\eeq
Using the mass ratios
\(
m_j
=
\frac{\sin\frac{j\pi}{h}}
{\sin\frac{\pi}{h}}m_1 \)
together with \eqref{Acoefficients}, the linear system
\eqref{rightConstraint}--\eqref{leftConstraint} reduces to a single equation for the combination entering \eqref{bulkIntegral}. The resulting bulk-energy coefficient is
\beq
\epsilon_{\rm bulk}
=
-\frac{m_1}{
8
\left(
4\cos^2\frac{\pi}{h}-1
\right)\sin\frac{\pi}{h}
},
\qquad
n\geq2.
\label{bulkCoefficient}
\eeq

\subsection{Leading perturbative correction}
\label{LOpert}

The leading perturbative correction can be extracted by writing
\beq
\epsilon_I=\epsilon_I^{\rm as}+\delta\epsilon_I,
\qquad
\epsilon_I^{\rm as}
=
\epsilon_I^K+\epsilon_I^A-\epsilon_I^0,
\eeq
and introducing
\beq
g_I
=
\log\left(1+e^{-\epsilon_I^{\rm as}}\right)
-
\left(L_I^K+L_I^A-L_I^0\right).
\eeq
Linearizing the TBA equations in the kink--antikink overlap gives an
integral equation for $\delta\epsilon_I$. As shown in
Appendix~\ref{appB}, the terms involving $\delta\epsilon_I$ cancel from
the correction to the scaling function, leaving the compact expression
\beq
\delta c(r)
=
\frac{6}{\pi^2}
\sum_{I=1}^{2n}
\infint d\theta\,
g_I^K(\theta)\,
\partial_\theta\epsilon_I^K(\theta).
\label{leadingCorrectionSummary}
\eeq

The volume dependence of $g_I^K$ follows from the plateau expansion of
the kink functions. Denoting
\(
y=2-2h_{\rm rel}
=
2/P_n,
\)
the periodicity
\(
Y_I(\theta+i\pi P_n)=Y_I(\theta)
\)
implies an expansion of the form
\beq
\epsilon_I^+(x)-\epsilon_I^0
=
\sum_{k=1}^{\infty}
a_{I,k}e^{kyx},
\qquad
x\to-\infty.
\label{kinkPlateauExpansion}
\eeq
Indeed, every term in \eqref{kinkPlateauExpansion} is invariant under
$x\to x+i\pi P_n$. By parity, the corresponding antikink expansion is
\beq
\epsilon_I^-(x)-\epsilon_I^0
=
\sum_{k=1}^{\infty}
a_{\bar I,k}e^{-kyx},
\qquad
x\to+\infty,
\eeq
where $\bar I$ denotes the particle with the same species label and
opposite chirality.

In the kink region, it is convenient to introduce the shifted rapidity
\(
x=\theta+\log ( r/2)
=\theta-\Lambda.
\)
Since the argument of the antikink function is then $x+2\Lambda$, its
leading plateau tail behaves as
\beq
\epsilon_I^A(\theta)-\epsilon_I^0
=
a_{\bar I,k_0}
\left(\frac{r}{2}\right)^{2k_0y}
e^{-yk_0x}
+
O\left(r^{2(k_0+1)y}\right).
\eeq
where $a_I,k_0$ is smallest non-vanishing coefficient.  
Consequently,
\beq
g_I^K(\theta)
=
\left(\frac{r}{2}\right)^{2k_0y}
g_I^+(x)
+
O\left(r^{2(k_0+1)y}\right),
\label{gKscaling}
\eeq
where the volume-independent shifted source is
\beq
g_I^+(x)
=
a_{\bar I,k_0}e^{-k_0yx}
\left[
\frac{1}{1+e^{\epsilon_I^0}}
-
\frac{1}{1+e^{\epsilon_I^+(x)}}
\right].
\label{gplusDefinition}
\eeq
Substitution into \eqref{leadingCorrectionSummary} therefore gives
\beq
\delta c(r)
=
\left(\frac{r}{2}\right)^{2k_0y}
\frac{6}{\pi^2}
\sum_{I=1}^{2n}
\infint dx\,
g_I^+(x)\,
\partial_x\epsilon_I^+(x)
+
O\left(r^{2(k_0+1)y}\right).
\label{leadingPerturbativeCoefficient}
\eeq
Thus the leading kink--antikink overlap is determined entirely by the
volume-independent kink solution and its leading plateau coefficient.

\subsection{Numerical analysis}

We tested the analytical results numerically for the first four theories, $n=1,2,3,4$, by solving the universal TBA equations. Independent implementations were developed in Python, using the \texttt{numpy} and \texttt{scipy} packages, and in Julia, the latter providing the higher-precision results quoted below. The numerical solution employed fast Fourier transforms and made direct use of the analytical Fourier-space form of the universal kernel \eqref{universalKernel}. The rapidity integrals were discretized on grids typically covering the interval $[-80,80]$, with between $2^{15}$ and $2^{18}$ grid points depending on the required accuracy. The iteration was terminated when the scaling function changed by less than $10^{-15}$ between successive steps.

\begin{figure}[t]
    \centering
    \includegraphics[width=0.72\linewidth]{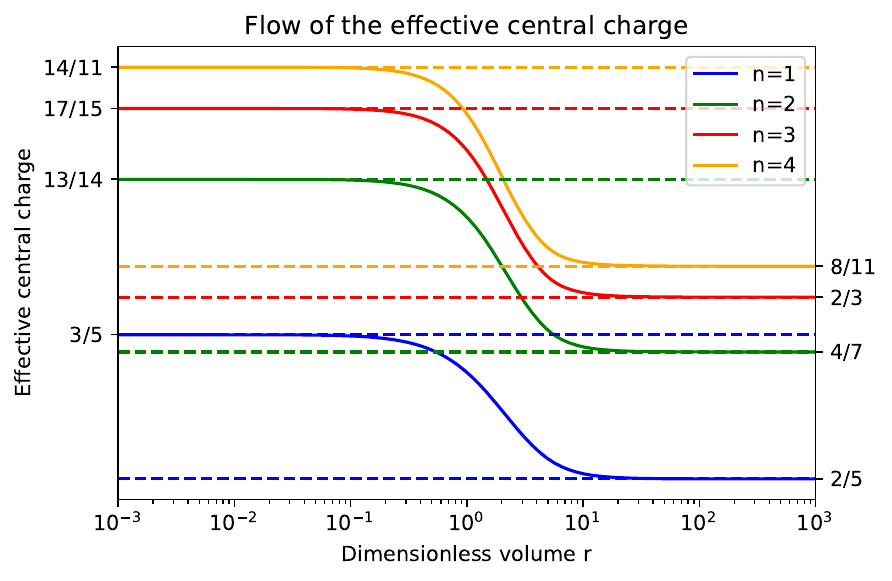}
    \caption{%
    Effective central charge \(c(r)\) obtained from the numerical TBA
    for \(n=1,2,3,4\), as a function of the dimensionless volume
    \(r=m_1 R\).  The dashed horizontal lines indicate the exact UV and
    IR effective central charges.  The numerical solutions interpolate
    between the corresponding conformal fixed points as \(r\) increases.
    }
    \label{fig:central-charge-flow}
\end{figure}

The full finite-volume solutions provide a direct numerical picture of
the RG trajectories.  Figure~\ref{fig:central-charge-flow} shows the
scaling function \(c(r)\) for \(n=1,2,3,4\).  In each case it
interpolates smoothly between the analytically predicted ultraviolet
and infrared effective central charges.

In addition to the finite-volume TBA, we solved the volume-independent kink equations. The UV effective central charge obtained from the kink integral \eqref{ceffUV} proved particularly stable numerically. The bulk-energy integral \eqref{bulkIntegral}, on the other hand, is considerably more sensitive to the exponentially small tails of the kink functions and was therefore less accurate. For this reason, the most precise numerical determination of the bulk term was obtained by fitting the small-volume expansion of the finite-volume TBA data. We also fitted the plateau expansion \eqref{kinkPlateauExpansion} of the kink pseudoenergies to determine the first non-vanishing exponent $k_0 y$. As discussed above, this exponent is $1$ for $n=1$, $5/4$ for $n=2$, $7/5$ for $n=3$ and $3/2$ for $n=4$.

For $n=1$, we reproduced the results of Ref.~\cite{RavaniniStanishkovTateo} for the bulk terms and for the first six perturbative coefficients. Among these, only $c_4$ and $c_6$ are non-vanishing. The same pattern persists for the higher-$n$ theories: both $c_2$ and the odd coefficients are compatible with zero within the numerical precision, although the accuracy naturally decreases at higher orders. Typical fitted magnitudes are
\[
c_1\sim10^{-12},
\qquad
c_2\sim10^{-10},
\qquad
c_3\sim10^{-8},
\qquad
c_5\sim10^{-5}.
\]
We therefore set these coefficients to zero in the final fits. Restricting the small-volume expansion to the non-vanishing even powers substantially improves the stability and accuracy of the fits. The fits were performed using the scaling function evaluated at $40$ volumes in the interval
\(
10^{-4}\leq r\leq10^{-2}.
\)

For $n=2$, after removing the odd coefficients, the fitted bulk-energy coefficient agrees with the analytical result to $\mathcal O(10^{-8})$. Setting also the numerically vanishing $c_2$ to zero gives
\beq
c_4=-1.397561438,
\qquad
c_6=0.9610049797.
\eeq
The value of $c_4$ agrees to seven significant figures with the independent numerical evaluation of the kink integral derived in Appendix~\ref{appB} for the leading non-vanishing perturbative correction.

For $n=3$, we find
\beq
c_4=-1.8823188581,
\qquad
c_6=2.22346503025.
\eeq
The value of $c_4$ obtained from the finite-volume fit agrees with the direct kink calculation to six significant figures.

For $n=4$, the first two non-vanishing coefficients are
\beq
c_4=-2.7825324485,
\qquad
c_6=4.7815572654,
\eeq
with the value of $c_4$ again confirmed independently from the kink solution to approximately six significant figures.

Altogether, the numerical analysis provides several independent checks of the proposed RG flows. The finite-volume TBA reproduces the expected UV and IR effective central charges, while the fitted bulk-energy coefficient agrees with the analytical result. The asymptotic behavior of the kink solutions, together with direct fits of the small-volume expansion, confirms the vanishing of the odd perturbative coefficients and of $c_2$. Finally, the first non-vanishing perturbative coefficient extracted from the volume dependence agrees with its independent expression in terms of the kink solution. These checks provide a stringent numerical test of both the identification of the UV fixed points and the perturbations generating the flows.

\subsection{Identification of the UV CFT and perturbation}

The analytical results above, together with the numerical small-volume
expansion, identify the ultraviolet fixed point as the higher-fusion-level
minimal model
\beq
\mathcal{M}({3,2n+3;n})
=
\frac{
\widehat{\mathfrak{su}}(2)_n
\oplus
\widehat{\mathfrak{su}}(2)_{-\frac{1}{2}}
}{
\widehat{\mathfrak{su}}(2)_{n-\frac{1}{2}}
}.
\label{UVcoset}
\eeq
Indeed, its effective central charge is
\beq
c_{\rm eff}
\left(
\mathcal{M}({3,2n+3;n})
\right)
=
\frac{3n}{n+2}
-
\frac{2n}{2n+3},
\eeq
in agreement with the TBA result \eqref{ceffUV}. The numerical expansion
provides further support for this identification: all odd perturbative
coefficients vanish, while $c_2=0$, so that the first nonzero correction
is proportional to $c_4$.

The periodicity \eqref{periodicity} fixes the chiral dimension of the
perturbing operator to
\beq
h_{\rm rel}
=
\frac{2n+7}{4n+8}.
\eeq
This value occurs in the spectrum of \(\mathcal{M}(3,2n+3;n)\). In the
parafermionic description, the corresponding primary field is
\beq
\Phi_{1,2,1}
=
\sigma_1\phi_{2,1},
\label{pertfield}
\eeq
whose conformal dimension is
\beq
h\left(\Phi_{1,2,1}\right)
=
\frac{
(4n+3)^2-(2n)^2
}{
12n(2n+3)
}
+
\frac{n-1}{2n(n+2)}
=
h_{\rm rel}.
\label{pertfielddimension}
\eeq
Here
\beq
h(\sigma_k)
=
\frac{k(n-k)}{2n(n+2)}
\eeq
is the conformal dimension of the spin field of the
\(\mathbb Z_n\) parafermion theory. For \(n=1\), the coset reduces to the
Virasoro minimal model \(\mathcal M(3,5)\), and the perturbation is
\(\phi_{2,1}\). For \(n=2\), it reduces to the superminimal model
\(S\mathcal M(3,7)\), and \(\Phi_{2,1}^{(1)}\) is the Ramond primary
\(\phi_{2,1}^{(\mathrm R)}\) of dimension \(11/16\).

We therefore identify the family of massless RG flows as
\beq
\mathcal{M}({3,2n+3;n})
+
\lambda\int d^2x\,\Phi_{1,2,1}
\quad\longrightarrow\quad
\mathcal{M}({2,2n+3}).
\label{mainRGflow}
\eeq

\subsection{Vanishing of the second-order coefficient}

The numerical TBA analysis indicates that all odd perturbative
coefficients vanish and that the second-order coefficient is also zero,
so that the first nonvanishing correction is $c_4$. The vanishing of
the odd orders follows from the fusion selection rule in the
$\widehat{\mathfrak{su}}(2)_n$ sector: the perturbing field carries
spin $1/2$, and an odd number of such fields cannot connect the
lowest-weight state back to itself.

The vanishing of $c_2$ is more subtle, since the relevant four-point
function is nonzero. Denoting the lowest-weight state of the UV theory
by $\Omega_n$ and the perturbing field by
$\Phi_n=\Phi_{1,2,1} $, its chiral dimension being
\(
h_n=h_{\mathrm{rel}},
\)
the second-order coefficient is proportional to
\beq
\mathcal I_n
=
\int_{\mathbb C}d^2z\,
|z|^{2h_n-2}
\left\langle
\Omega_n(\infty)\Phi_n(1)
\Phi_n(z,\bar z)\Omega_n(0)
\right\rangle .
\label{c2integral}
\eeq
Using the $\beta\gamma$-ghost realization of
$\widehat{\mathfrak{su}}(2)_{-1/2}$, the coset conformal block reduces
to
\beq
\mathcal F_n(z)
=
z^{-1/2}(1-z)^{-2h_n}(1+z),
\label{genericC2block}
\eeq
and therefore
\beq
\left\langle
\Omega_n(\infty)\Phi_n(1)
\Phi_n(z,\bar z)\Omega_n(0)
\right\rangle
=
\mathcal N_n
|z|^{-1}|1-z|^{-4h_n}|1+z|^2 .
\label{genericC2correlator}
\eeq
After analytic regularization, the integral \eqref{c2integral}
vanishes identically. This generalizes the direct Virasoro calculation
for $n=1$ \cite{RavaniniStanishkovTateo} and the corresponding
Ramond-sector calculation in the superminimal model
$S\mathcal M(3,7)$ at $n=2$. Thus
\(
c_2=0
\)
for the complete family, in agreement with the numerical TBA
expansion.

\subsection{Other mixed-chirality solutions and ultraviolet completeness}
\label{subsec:othersolutions}

The bootstrap equations admit a larger class of pole-free
mixed-chirality amplitudes than the minimal solution studied above. It
is important, however, to distinguish the existence of an algebraically
consistent scattering theory from the existence of a conformal
ultraviolet limit.

The elementary solutions can be obtained by folding the corresponding
\(A_{2n}^{(1)}\) solutions. In the folded theory they may be labelled by
their fundamental amplitudes
\beq
S_{11}^{(k)}(\theta)=[-k]_\theta,
\qquad
k=1,\ldots,n,
\label{elementarySolutions}
\eeq
with all remaining \(S_{ab}^{(k)}\) determined by the bootstrap.
Products of these elementary amplitudes provide further solutions. The
first member, \(k=1\), is the minimal solution analyzed in detail in
this work, while \(k=2\) gives the diagonal solution
\beq
S_{ab}^{(2)}(\theta)=S_{ab}^{-1}(\theta).
\label{diagonalSolution}
\eeq
The solution obtained by squaring the minimal amplitude,
\beq
S_{ab}^{\rm sat}(\theta)=
\left(S_{ab}^{(1)}(\theta)\right)^2 ,
\label{saturatedSolution}
\eeq
is the folded counterpart of the saturated solution of the
\(A_{2n}^{(1)}\) classification.

The ultraviolet behavior provides a much stronger restriction. For the
minimal solution the constant \(Y\)-system possesses the positive
solution \eqref{UVplateaus}, leading to the ultraviolet CFT identified
above. The diagonal solution also has a regular constant solution. In
the central plateau the same- and mixed-chirality kernels cancel,
giving
\beq
Y_a^{\rm R}=Y_a^{\rm L}=1,
\qquad a=1,\ldots,n.
\label{diagonalPlateau}
\eeq
At the outer ends of the kink solution one recovers the massive
\(A_{2n}^{(2)}\) plateau. The dilogarithm sum then gives
\beq
c_{\rm eff}^{\rm UV,diag}
=
n-c_{\rm eff}^{\rm IR}=
\frac{n(2n+1)}{2n+3}.
\label{diagonalCentralCharge}
\eeq
The corresponding \(Y\)-systems for \(n=1,2,3\) have period
\(4\pi i\), or equivalently a half-period \(2\pi i\) accompanied by
the interchange of left- and right-movers. This is consistent with a
perturbing chiral dimension
\(
h_{\rm diag}=\frac34 .
\label{diagonalDimension}
\)
For \(n>1\), however, we have not identified a conformal field theory
having simultaneously the effective central charge
\eqref{diagonalCentralCharge} and the required operator content.

The situation changes for the saturated solution. Although
\eqref{saturatedSolution} satisfies all algebraic bootstrap,
crossing-unitarity and analyticity requirements, numerical analysis of
its constant \(Y\)-system does not reveal a positive real solution.
Consequently the corresponding TBA does not develop the finite
plateau required for a conventional ultraviolet conformal limit.
This provides an explicit example in which a perfectly consistent
mixed-chirality bootstrap solution does not define a UV-complete
massless RG flow.

Preliminary investigations of the remaining higher solutions indicate
the same tendency. We therefore conjecture that, within the folded
\(A_{2n}^{(2)}\) family, the minimal and diagonal amplitudes are the
only elementary solutions admitting regular conformal ultraviolet
limits, while the higher solutions encounter a singularity before the
UV regime is reached. A systematic analysis of these solutions will be
presented elsewhere.

\newpage
\section{Non-invertible symmetries and Verlinde defect lines}
In this section we investigate the non-invertible symmetries preserved by the perturbation. 

\subsection{Preserved Verlinde lines along RG flows}
The fusion algebra of Verlinde defect lines is closely related to the representation theory of quantum groups at roots of unity. In particular, the truncated tensor products of admissible quantum-group representations reproduce the fusion rules of rational conformal field theories, with the appropriate selection rules and field identifications imposed in coset constructions \cite{AlvGomSie}.
This provides a connection with the quantum-group restriction of the sine-Gordon model discussed in sect.~2.1. 
There is, however, an important distinction between the two constructions: the quantum-group symmetry underlying the scattering theory and the topological defect operators acting on CFT states are not identical notions. 
In particular, the preservation of a Verlinde line along an RG flow requires that it commute with the perturbing operator.
For the family of defects considered below, this condition will follow from the fact that their eigenvalues depend only on the denominator representation label, which remains unchanged under fusion with the perturbing field.

Let \(\mathbb D_a\) be a topological defect line in a CFT deformed by a
relevant primary field \(\phi_b\). The defect is preserved by the
deformation if it commutes with the perturbing operator on the
cylinder,
\beq
\mathbb D_a\phi_b\vert\Phi\rangle
=
\phi_b \mathbb D_a\vert\Phi\rangle ,
\label{vdlcomm}
\eeq
for every state \(\vert\Phi\rangle\). Since a topological defect
commutes with the chiral algebra, it is sufficient to verify this
condition on the primary states.

For a Verlinde defect, the action on a primary state is diagonal:
\beq
\mathbb D_a\vert\phi_c\rangle
=
\ell_a(c)\vert\phi_c\rangle,
\qquad
\ell_a(c)=\frac{S_{ac}}{S_{0c}},
\label{moddim}
\eeq
where \(0\) denotes the identity primary. Using
\[
\phi_b\times\phi_c
=
\sum_d N_{bc}^{\phantom{bc}d}\phi_d,
\]
the commutation condition \eqref{vdlcomm} becomes
\beq
N_{bc}^{\phantom{bc}d}\neq0
\qquad\Longrightarrow\qquad
\ell_a(d)=\ell_a(c).
\label{niscond}
\eeq
In particular, applying this condition to the vacuum gives the
necessary relation
\beq
\ell_a(b)=\ell_a(0),
\qquad\text{or equivalently}\qquad
\frac{S_{ab}S_{00}}{S_{0b}S_{a0}}=1.
\label{vacuumcondition}
\eeq
In a non-unitary theory this vacuum condition need not be sufficient,
and the full condition \eqref{niscond} should in general be checked.

If \eqref{vdlcomm} holds, the defect can be moved through every
insertion of the perturbing operator and therefore remains topological
in conformal perturbation theory. Its fusion rules, quantum dimension
\[
d_a=\ell_a(0)=\frac{S_{a0}}{S_{00}},
\]
and other categorical data then provide RG invariants that must be
compatible with those of the infrared theory.

For the Virasoro minimal model \(\mathcal M(p,q)\), in the conventions
used here, the primary-field fusion rules are
\beq
\phi_{r,s}\times\phi_{m,\ell}
=
\sum_{r'\in\Lambda^{p}_{r,m}}
\sum_{s'\in\Lambda^{q}_{s,\ell}}
\phi_{r',s'},
\label{minimalfusion}
\eeq
where
\beq
\Lambda^\alpha_{\beta,\gamma}
=
\left\{
|\beta-\gamma|+1,\,
|\beta-\gamma|+3,\,
\ldots,\,
\min\left(
\beta+\gamma-1,\,
2\alpha-1-\beta-\gamma
\right)
\right\}.
\label{fusionruleset}
\eeq
This criterion was used in Ref.~\cite{NakTan} to identify Verlinde
defects preserved by particular relevant perturbations and to
constrain possible RG flows between Virasoro minimal models. Such
matching conditions provide powerful selection rules, although by
themselves they do not prove the existence of the corresponding flow.

\subsection{Preserved defects in the higher-fusion-level minimal models}

We now apply this construction to the coset theories
\beq
\mathcal{M}(t_1,t_2;n)
=
\frac{
\widehat{\mathfrak{su}}(2)_n
\oplus
\widehat{\mathfrak{su}}(2)_{p-2}
}{
\widehat{\mathfrak{su}}(2)_{n+p-2}
},\qquad p=\frac{nt_1}{t_2-t_1},
\label{nLcoset}
\eeq
as defined in \eqref{higherminimalcoset}.
We denote their primary fields by
\(\Phi_{j,r,s}\), where \(j\), \(r\), and \(s\) label the
representations of the three affine factors. Suppressing the
admissibility labels fixed by the coset selection rules, the modular
\(S\)-matrix factorizes as
\beq
\mathbb S_{(j,r,s),(j',r',s')}
=
S^{[n]}_{j+1,j'+1}\,
S^{[t_1]}_{r,r'}\,
\left(S^{[t_2]}_{s,s'}\right)^* .
\label{cosetS}
\eeq
For fractional \(p\), the factors in \eqref{cosetS} are understood as
the admissible-level modular matrices \cite{KohSorba}.

Consider the Verlinde defects
\beq
\mathbb D_N
\equiv
\mathbb D_{(0,1,N)}.
\label{defDN}
\eeq
In the following we choose representatives in the admissibility sector \(k=0\) of Eqs.\eqref{J1J2}-\eqref{nonunitaryselection}. 
For the defects \(D_N=D_{(0,1,N)}\), the coset selection rule then requires \(N\) to be odd, so that we may take
\(N=1,3,\ldots,2n+1\), modulo the corresponding field identifications.
Their action on an arbitrary coset primary is
\bea
\mathbb D_N\vert\Phi_{j,r,s}\rangle
&=&
\lambda_N(s)\vert\Phi_{j,r,s}\rangle,
\nonumber\\
\lambda_N(s)
&=&
\frac{
\mathbb S_{(0,1,N),(j,r,s)}
}{
\mathbb S_{(0,1,1),(j,r,s)}
}
=
\frac{
S^{[t_2]}_{N,s}
}{
S^{[t_2]}_{1,s}
},
\label{defecteigenvalue}
\eea
where we used that the modular $S$-matrix elements are real.
The important point is that \(\lambda_N(s)\) depends only on the
denominator label \(s\).

The coset fusion rules take the form
\beq
\Phi_{j,r,s}\times\Phi_{j',r',s'}
=
\sum_{j''+1\in\Lambda^{n+2}_{j+1,j+1}}
\sum_{r''\in\Lambda^{t_1}_{r,r'}}
\sum_{s''\in\Lambda^{t_2}_{s,s'}}
\Phi_{j'',r'',s''},
\label{cosetfusion}
\eeq
in terms of the sets defined in \eqref{fusionruleset}.
For the perturbing field \(\Phi^{(1)}_{2,1}\), this reduces to
\beq
\Phi_{1,2,1}\times\Phi_{j,r,s}
=
\sum_{j''+1\in\Lambda^{n+2}_{2,j+1}}
\sum_{r''\in\Lambda^{t_1}_{2,r}}
\Phi_{j'',r'',s}.
\label{perturbingfusion}
\eeq
In particular, the denominator label \(s\) is unchanged in every
fusion channel. Equation \eqref{defecteigenvalue} therefore implies
\beq
\lambda_N(j'',r'',s)
=
\lambda_N(j,r,s)
\eeq
for every primary appearing on the right-hand side of
\eqref{perturbingfusion}. Hence
\beq
\mathbb D_N\,
\Phi_{1,2,1}
\vert\Phi_{j,r,s}\rangle
=
\Phi_{1,2,1}\,
\mathbb D_N
\vert\Phi_{j,r,s}\rangle
\label{commutedefect}
\eeq
for all \(j,r,s\). The complete family of defects \(\mathbb D_N\) is
therefore preserved by the perturbation.

For the UV theories relevant to the present flows, \(p-2=-1/2\), and the
quantum dimensions of these defects are
\bea
d_N^{\rm UV}
=
\lambda_N(1)
=
\frac{
S^{[t_2]}_{N,1}
}{
S^{[t_2]}_{1,1}
}
=
\frac{
\sin\frac{2\pi N}{2n+3}
}{
\sin\frac{2\pi}{2n+3}
}.
\label{UVquantumdimension}
\eea
On the infrared side, the Verlinde line \(\mathbb D_{(1,N)}\) of
\(\mathcal M(2,2n+3)\) has, in the same modular-\(S\) convention,
\beq
d_{(1,N)}^{\rm IR}
=
\frac{
S^{\rm IR}_{(1,N),(1,1)}
}{
S^{\rm IR}_{(1,1),(1,1)}
}
=
\frac{
\sin\frac{2\pi N}{2n+3}
}{
\sin\frac{2\pi}{2n+3}
}.
\label{IRquantumdimension}
\eeq
Thus,
\(
d_N^{\rm UV}=d_{(1,N)}^{\rm IR},
\)
for the allowed values of \(N\), modulo the corresponding coset and
Kac-table identifications. This matching is consistent with the
identification
\beq
\mathbb D_{(0,1,N)}^{\rm UV}
\quad\longrightarrow\quad
\mathbb D_{(1,N)}^{\rm IR}
\label{defectflow}
\eeq
along the RG flow for all allowed $N$. The preserved Verlinde defects therefore provide an
independent non-perturbative consistency check of the flows obtained
from the exact massless scattering theory and the TBA analysis.

\section{Conclusions}
\label{sec:conclusions}

We have constructed an infinite family of exact massless factorized
scattering theories describing RG flows between non-unitary conformal
fixed points. Starting from the \(A_{2n}^{(2)}\) massive scattering
theory associated with the \(\phi_{1,3}\) perturbation of
\(\mathcal M(2,2n+3)\), we retained its same-chirality amplitudes in
the massless limit and selected the minimal pole-free solution of the
mixed-chirality bootstrap equations. The resulting thermodynamic Bethe
ansatz admits a universal formulation in terms of two tadpole systems
coupled node by node.

The ultraviolet plateau solution and the periodicity of the associated
\(Y\)-system give
\beq
c_{\mathrm{eff}}^{\mathrm{UV}}
=
\frac{3n}{n+2}
-
\frac{2n}{2n+3},
\qquad
h_{\mathrm{rel}}
=
\frac{2n+7}{4n+8}.
\eeq
These quantities identify the RG trajectory as
\beq
\mathcal M(3,2n+3;n)
+
\lambda\int d^2x\,\Phi_{2,1}^{(1)}
\quad\longrightarrow\quad
\mathcal M(2,2n+3).
\eeq
For \(n=1\), this produces the flow from
\(\mathcal M(3,5)\) to the Yang--Lee CFT \(\mathcal M(2,5)\) generated by \(\phi_{2,1}\).
It is noticeable that this is not one of the non-unitary Zamolodchikov 
flows \eqref{zamRGflow}, neither belongs to the recent RG flows generated by the non-invertible symmetries found in \cite{NakTan}. 
For \(n=2\),
the ultraviolet theory is the superminimal model
\(S\mathcal M(3,7)\) perturbed by its Ramond primary
\(\phi_{2,1}^{(\mathrm R)}\).

The identification is supported by several independent checks. The
finite-volume TBA reproduces the ultraviolet and infrared effective
central charges and the analytically determined bulk contribution.
The fusion rules imply the absence of all odd perturbative orders,
whereas the second-order coefficient vanishes through an analytic
cancellation in the four-point integral. Consequently, the first
nonzero perturbative term is \(c_4\). For the theories studied
numerically, its value extracted from the finite-volume scaling
function agrees with the independent expression obtained from the
volume-independent kink solution.

We have also identified a family of Verlinde defect lines that
commutes with the perturbing field. Their modular-\(S\) eigenvalues
and quantum dimensions match at the two conformal endpoints,
consistently with
\beq
\mathbb D_N^{\mathrm{UV}}
\quad\longrightarrow\quad
\mathbb D_{(1,N)}^{\mathrm{IR}}.
\eeq
The scattering bootstrap, TBA, conformal perturbation theory, and
defect analysis therefore provide mutually consistent descriptions of
the same family of non-unitary massless RG flows.

The bootstrap admits several further pole-free mixed-chirality
solutions, but our analysis indicates that ultraviolet completeness is
a substantially stronger condition than bootstrap consistency. Besides
the minimal solution studied in detail here, the diagonal amplitude
\(S^{\rm RL}=S^{-1}\) possesses a regular ultraviolet plateau with
\beq
c_{\rm eff}^{\rm UV}
=
\frac{n(2n+1)}{2n+3},
\eeq
although we have not yet identified the corresponding UV CFT for
general \(n\). By contrast, for the saturated solution we find no
positive real solution of the constant \(Y\)-system, indicating the
absence of a conventional conformal UV limit. This suggests that the
requirement of a regular ultraviolet completion may select only a
small subset of the algebraically allowed left--right scattering
amplitudes. Establishing this classification for the remaining
bootstrap solutions and identifying the UV theory associated with the
diagonal solution are interesting problems for future work.

The present construction shows that the massless bootstrap can be used
systematically to obtain ultraviolet completions of terminal
non-unitary minimal models. Natural extensions include determining the
exact mass--coupling relation, constructing excited-state TBA systems,
and analyzing the diagonal and saturated mixed-chirality solutions and
the corresponding defect data.

\section*{Acknowledgements}

CA thanks a hospitality of Wigner Research Center where this work has been initiated. This work was supported by the National Research Foundation of Korea (NRF) grant
RS-2026-25472596 (CA) and the NKKP-Advanced 152467 grant (ZB, ES) of the Hungarian National Research, Development and Innovation Office. 

\appendix
\renewcommand{\theequation}{A.\arabic{equation}}
\setcounter{equation}{0}
\section{Higher-fusion-level minimal CFTs}
\label{appA}

In this appendix, we summarize the diagonal coset construction of the
higher-fusion-level minimal CFTs and specialize it to the non-unitary
models relevant to the flows studied in the main text.

\subsection{Unitary diagonal cosets}

Consider the diagonal $su(2)$ coset CFTs
\beq
\mathcal{M}(p,p+n;n)
\equiv
\frac{
\widehat{\mathfrak{su}}(2)_n
\oplus
\widehat{\mathfrak{su}}(2)_{p-2}
}{
\widehat{\mathfrak{su}}(2)_{n+p-2}
},
\label{unitarycoset}
\eeq
where \(n\in\mathbb Z_{\geq 1},p\in\mathbb Z_{\geq 3}\). 
Using the central charge of the affine WZW CFT
\(\widehat{\mathfrak{su}}(2)_k\)
\beq
c_k=\frac{3k}{k+2},
\eeq
we can find those of the coset CFTs
\beq
c_{n,p}
=
\frac{3n}{n+2}
-
\frac{6n}{p(p+n)}.
\label{unitaryhigherc}
\eeq
For \(n=1\), the coset reduces to the unitary Virasoro minimal model
\beq
\mathcal{M}(p,p+1;1)=\mathcal M(p,p+1).
\eeq
The case \(n=2\) gives the unitary \(N=1\) superminimal series, while
\(n\geq3\) gives higher-fusion-level minimal models with extended
\(\mathbb Z_n\)-parafermionic symmetry.

An integrable highest-weight representation of
\(\widehat{\mathfrak{su}}(2)_n\) is labelled by
\(\ell=0,\ldots,n\), with conformal weight
\beq
h_\ell^{(n)}
=
\frac{\ell(\ell+2)}{4(n+2)}.
\label{wzwhws}
\eeq
Then, the coset primaries can be represented by $\Phi_{\ell,r,s}$ where
\beq
0\leq\ell\leq n,
\qquad
1\leq r\leq p-1,
\qquad
1\leq s\leq p+n-1,
\eeq
subject to the coset selection rules and field identifications.

It is useful to separate their conformal weights into a generalized
Kac contribution and a parafermionic contribution:
\beq
h_{\ell,r,s}
=
\frac{
\bigl(r(p+n)-sp\bigr)^2-n^2
}{
4np(p+n)
}
+
h_{\ell,m}^{(n)}
+
N_{\ell,r,s},
\label{unitaryweights}
\eeq
where
\beq
h_{\ell,m}^{(n)}
=
\frac{\ell(\ell+2)}{4(n+2)}
-
\frac{m^2}{4n}
\qquad
(\mathrm{mod}\ \mathbb Z),
\label{PFweight}
\eeq
is the \(\mathbb Z_n\)-parafermionic weight. The charge \(m\) obeys
\beq
m\equiv r-s\pmod{2n},
\qquad
\ell-m\in2\mathbb Z,
\label{PFselection}
\eeq
and \(N_{\ell,r,s}\) is the grade at which the corresponding
denominator representation occurs.

For the spin-field branch \(m=\pm\ell\), with vanishing grade shift,
\eqref{PFweight} reduces to
\beq
h(\sigma_\ell)
=
\frac{\ell(n-\ell)}{2n(n+2)}.
\label{spinfieldweight}
\eeq
Consequently,
\beq
h_{\ell,r,s}
=
\frac{
\bigl(r(p+n)-sp\bigr)^2-n^2
}{
4np(p+n)
}
+
\frac{\ell(n-\ell)}{2n(n+2)}
\label{unitaryspinbranch}
\eeq
on this branch.

For \(n=2\), the sectors \(\ell=0,2\) are Neveu--Schwarz, while
\(\ell=1\) is Ramond. The superconformal primary weights may be written
as
\beq
h_{r,s}
=
\frac{
\bigl(r(p+2)-sp\bigr)^2-4
}{
8p(p+2)
}
+
\frac{1}{16}\,
\delta_{r+s,\mathrm{odd}}.
\label{superdim}
\eeq
For general \(n\), the sectors labelled by \(\ell\) are higher-level
analogues of the Neveu--Schwarz and Ramond sectors, modulo the standard
field identifications.

\subsection{Non-unitary diagonal cosets}

We now allow the second affine level in \eqref{unitarycoset} to be fractional:
\beq
p=\frac{t_1}{u},
\qquad
\gcd(t_1,u)=1,
\qquad
t_1\geq2,
\qquad
u\geq1.
\label{admissiblelevel}
\eeq
The denominator level satisfies
\beq
n+p=\frac{t_2}{u},
\qquad
t_2=t_1+nu.
\label{t2definition}
\eeq
An admissible highest weight of
\(\widehat{\mathfrak{su}}(2)_{p-2}\) can be parametrized as
\beq
J=j-kp,
\qquad
1\leq j\leq t_1-1,
\qquad
0\leq k\leq u-1,
\label{admissibleweight}
\eeq
with conformal weight
\beq
h_J^{(p-2)}
=
\frac{J^2-1}{4p}.
\label{wzwAhws}
\eeq

In terms of this parametrization, the coset CFT $\mathcal{M}(p,p+n;n)$ with
\beq
p=\frac{nt_1}{t_2-t_1},
\label{fractionallevelMM}
\eeq
defines
the higher-fusion-level minimal model \(\mathcal M(t_1,t_2;n)\) 
\beq
\mathcal M(t_1,t_2;n)
\equiv
\frac{
\widehat{\mathfrak{su}}(2)_n
\oplus
\widehat{\mathfrak{su}}(2)_{p-2}
}{
\widehat{\mathfrak{su}}(2)_{n+p-2}
}
\label{higherminimalcoset}
\eeq
with
the central charge in \eqref{unitaryhigherc} 
\beq
c_{t_1,t_2;n}
=
\frac{3n}{n+2}
-
\frac{6(t_2-t_1)^2}{nt_1t_2}.
\label{higherminimalc}
\eeq
The cases \(n=1\) and \(n=2\) reproduce the Virasoro and
\(N=1\) superconformal minimal models, respectively.

The admissible numerator and denominator weights may be written as
\beq
J_1=r-k\frac{t_1}{u},
\qquad
J_2=s-k\frac{t_2}{u},
\label{J1J2}
\eeq
where
\beq
1\leq r\leq t_1-1,
\qquad
1\leq s\leq t_2-1,
\qquad
0\leq k\leq u-1.
\eeq
The associated parafermionic charge is
\beq
m\equiv J_1-J_2
\equiv r-s+nk
\pmod{2n},
\qquad
\ell-m\in2\mathbb Z.
\label{nonunitaryselection}
\eeq

The conformal weights take the form
\beq
h_{\ell,r,s}
=
\frac{
(t_2r-t_1s)^2-(t_2-t_1)^2
}{
4nt_1t_2
}
+
h_{\ell,m}^{(n)}
+
N_{\ell,r,s},
\label{generalnonunitaryweight}
\eeq
where \(h_{\ell,m}^{(n)}\) is given in \eqref{PFweight}. The grade
\(N_{r,s,\ell}\) and the allowed values of \(m\) are determined by the
branching functions. For the spin-field branch \(m=\pm\ell\), with
vanishing grade shift, this simplifies to
\beq
h_{\ell,r,s}
=
\frac{
(t_2r-t_1s)^2-(t_2-t_1)^2
}{
4nt_1t_2
}
+
\frac{\ell(n-\ell)}{2n(n+2)}.
\label{cosetdimnun}
\eeq
We denote the corresponding fields by
\beq
\Phi_{\ell,r,s}
\equiv
\sigma_\ell\phi_{r,s}.
\label{cosetprimaryform}
\eeq

\subsection{The models \(\mathcal M(3,2n+3;n)\) }

For the UV theories appearing in the main text,
we choose
\beq
p=\frac32,
\qquad
t_1=3,
\qquad
u=2,
\qquad
t_2=2n+3
\eeq
so that 
\beq
\mathcal M(3,2n+3;n)
=
\frac{
\widehat{\mathfrak{su}}(2)_n
\oplus
\widehat{\mathfrak{su}}(2)_{-\frac12}
}{
\widehat{\mathfrak{su}}(2)_{n-\frac12}
},
\label{ourcoset}
\eeq
with the central charge
\beq
c_{3,2n+3;n}
=
\frac{3n}{n+2}
-
\frac{8n}{2n+3}.
\label{ourcentralcharge}
\eeq

A representative of the lowest-dimensional primary is
\(\Phi_{1,n+1}^{(0)}\), equivalently
\(\Phi_{2,n+2}^{(0)}\) under the coset field identifications. Its
conformal dimension is
\beq
h_{\min}
=
-\frac{n}{4(2n+3)}.
\label{hmincoset}
\eeq
The effective central charge is therefore
\bea
c_{\mathrm{eff}}=
c_{3,2n+3;n}-24h_{\min}
=
\frac{3n}{n+2}
-
\frac{2n}{2n+3},
\label{ourceff}
\eea
in agreement with the ultraviolet value obtained from the TBA.

The relevant perturbing field is
\beq
\Phi_{1,2,1}
=
\sigma_1\phi_{2,1}.
\label{ourperturbation}
\eeq
Using \eqref{cosetdimnun}, its chiral conformal dimension is
\bea
h\left(\Phi_{1,2,1}\right)=
\frac{
(4n+3)^2-(2n)^2
}{
12n(2n+3)
}
+
\frac{n-1}{2n(n+2)}
=
\frac{2n+7}{4n+8}.
\label{ourperturbingdimension}
\eea
This coincides with the value inferred from the periodicity of the
\(Y\)-system. For \(n=1\), the theory reduces to
\(\mathcal M(3,5)\) perturbed by \(\phi_{2,1}\); for \(n=2\), it
reduces to the superminimal model \(S\mathcal M(3,7)\) perturbed by the
Ramond primary \(\phi_{2,1}^{(\mathrm R)}\).

\renewcommand{\theequation}{B.\arabic{equation}}
\setcounter{equation}{0}

\section{Leading perturbative correction to the ground-state energy}
\label{appB}

The subleading correction to the scaling function can be written as
\beq
\delta c(r)
=
\frac{3}{\pi^2}
\sum_{I=1}^{2n}
\infint d\theta\,d_I\,\delta L_I
=
\frac{6}{\pi^2}
\sum_{i=1}^{n}
\infint d\theta\,d_i\,\delta L_i,
\label{deltacdef}
\eeq
where in the second equality we used the symmetry between the right- and left-moving contributions.

We introduce the asymptotic functions
\bea
\epsilon_I^{\rm as}
&=&
\epsilon_I^K+\epsilon_I^A-\epsilon_I^0,
\nonumber\\
L_I^{\rm as}
&=&
L_I^K+L_I^A-L_I^0,
\eea
and define the corrections by
\beq
\epsilon_I=\epsilon_I^{\rm as}+\delta\epsilon_I,
\qquad
L_I=L_I^{\rm as}+\delta L_I.
\label{asymptoticdecomposition}
\eeq
Notice that $L_I^{\rm as}$ is defined additively and is therefore not, in general, equal to
$\log(1+e^{-\epsilon_I^{\rm as}})$.

Subtracting the kink, antikink, and plateau TBA equations from the full TBA gives
\bea
\delta\epsilon_I
&=&
\epsilon_I-\epsilon_I^K-\epsilon_I^A+\epsilon_I^0
\nonumber\\
&=&
-\varphi_{IJ}\star
\left(
L_J-L_J^K-L_J^A+L_J^0
\right),
\label{deltaepequation}
\eea
where the driving terms cancel. It is convenient to define
\beq
g_I
:=
\log\left(1+e^{-\epsilon_I^{\rm as}}\right)
-
L_I^{\rm as}.
\label{defg}
\eeq
Expanding $L_I$ around $\epsilon_I^{\rm as}$, we obtain
\beq
\delta L_I
=
g_I
-
\frac{\delta\epsilon_I}{1+e^{\epsilon_I^{\rm as}}}
+
O\left((\delta\epsilon_I)^2\right).
\label{deltaLexpansion}
\eeq
Equation \eqref{deltaepequation} therefore reduces, at linear order, to
\beq
\delta\epsilon_I
=
-\varphi_{IJ}\star g_J
+
\varphi_{IJ}\star
\frac{\delta\epsilon_J}
{1+e^{\epsilon_J^{\rm as}}}.
\label{linearizedequation}
\eeq

We now focus on the kink region, where
\[
\epsilon_I^A-\epsilon_I^0
\]
is exponentially small. Expanding \eqref{defg} around the plateau gives
\bea
g_I^K
&=&
\log\left(
1+e^{-\epsilon_I^K-\epsilon_I^A+\epsilon_I^0}
\right)
-
L_I^K
-
L_I^A
+
L_I^0
\nonumber\\
&=&
\left[
\frac{1}{1+e^{\epsilon_I^0}}
-
\frac{1}{1+e^{\epsilon_I^K}}
\right]
\left(
\epsilon_I^A-\epsilon_I^0
\right)
+
O\left(
(\epsilon_I^A-\epsilon_I^0)^2
\right).
\label{gkink}
\eea
The corresponding linearized kink equation is
\beq
\delta\epsilon_I^K
=
-\varphi_{IJ}\star
\left(
g_J^K
-
\frac{\delta\epsilon_J^K}
{1+e^{\epsilon_J^K}}
\right).
\label{linearkink}
\eeq

At leading perturbative order, the correction to the scaling function is obtained from the kink--antikink overlap. Using left--right symmetry, it is sufficient to consider the kink region:
\beq
\delta c(r)
=
\frac{6}{\pi^2}
\sum_{i=1}^{n}
\infint d\theta\,
d_i^K
\left(
g_i^K
-
\frac{\delta\epsilon_i^K}
{1+e^{\epsilon_i^K}}
\right).
\label{deltackink}
\eeq
Since $d_I^K=0$ for $I>n$, the sum can equivalently be extended to all multi-indices. Differentiating the kink TBA equations gives
\beq
d_I^K
=
\partial_\theta\epsilon_I^K
+
\varphi_{IJ}\star\partial_\theta L_J^K.
\label{drivingidentity}
\eeq
Substituting \eqref{drivingidentity} into \eqref{deltackink}, using the symmetry of the kernel and the linear equation \eqref{linearkink}, we find
\bea
\delta c(r)
&=&
\frac{6}{\pi^2}
\sum_{I=1}^{2n}
\infint d\theta\,
\left[
g_I^K\partial_\theta\epsilon_I^K
-
\frac{
\delta\epsilon_I^K\partial_\theta\epsilon_I^K
}{
1+e^{\epsilon_I^K}
}
-
\delta\epsilon_I^K\partial_\theta L_I^K
\right].
\eea
The last two terms cancel because
\beq
\partial_\theta L_I^K
=
-\frac{
\partial_\theta\epsilon_I^K
}{
1+e^{\epsilon_I^K}
}.
\eeq
We therefore arrive at the compact result
\beq
\delta c(r)
=
\frac{6}{\pi^2}
\sum_{I=1}^{2n}
\infint d\theta\,
g_I^K(\theta)\,
\partial_\theta\epsilon_I^K(\theta).
\label{sublfinalresult}
\eeq

\begin{figure}[h]
    \centering
    \includegraphics[width=0.5\linewidth]{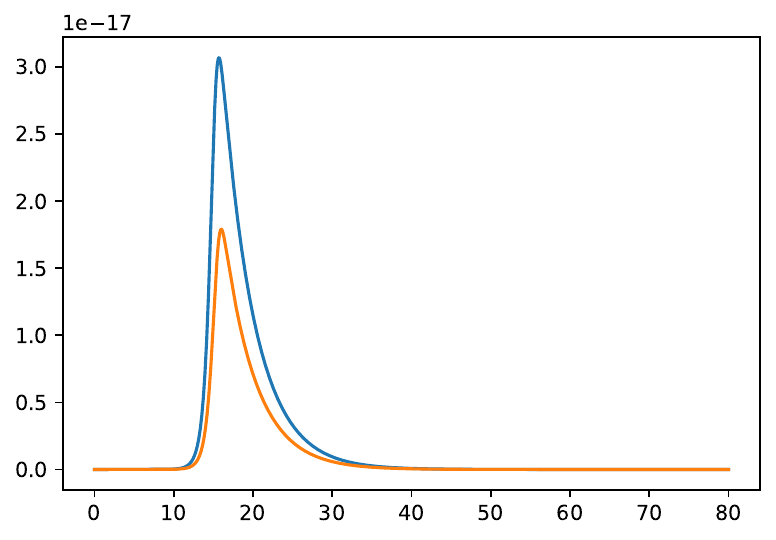}
    \caption{Integrands appearing in \eqref{sublfinalresult} as functions of rapidity. They are localized in the kink region and vanish rapidly towards negative rapidities.}
    \label{fig:integrand}
\end{figure}

\section{Second-order perturbative coefficient}
\label{app:c2vanishing}

We give here the conformal-perturbative derivation of the result
\(
c_2=0
\)
for the flows generated by
\beq
\Phi_n\equiv\Phi_{1,2,1},
\qquad
h_n=\frac{2n+7}{4n+8}.
\eeq
Let $\Omega_n$ denote the primary state of lowest conformal dimension
in the ultraviolet CFT, normalized such that
\(
\langle\Omega_n(\infty)\Omega_n(0)\rangle=1.
\)
Because the ground state of a non-unitary CFT is represented by
$\Omega_n$, rather than by the conformal vacuum, the second-order
ground-state energy correction involves the four-point function
\beq
G_n(z,\bar z)
=
\left\langle
\Omega_n(\infty)\Phi_n(1)
\Phi_n(z,\bar z)\Omega_n(0)
\right\rangle .
\label{defGn}
\eeq
Up to an overall normalization and the mass--coupling relation, the
coefficient $c_2$ is proportional to
\beq
\mathcal I_n
=
\int_{\mathbb C}d^2z\,
|z|^{2h_n-2}G_n(z,\bar z).
\label{defIn}
\eeq
The integral is understood by analytic continuation from a domain in
which all short-distance integrals converge.

All odd perturbative orders vanish by a fusion selection rule. In the
coset representation
\beq
\frac{
\widehat{\mathfrak{su}}(2)_n
\oplus
\widehat{\mathfrak{su}}(2)_{-1/2}
}{
\widehat{\mathfrak{su}}(2)_{n-1/2}
},
\eeq
the perturbing operator carries spin $1/2$ in the
$\widehat{\mathfrak{su}}(2)_n$ factor. An odd number of perturbing
fields therefore cannot fuse to the representation required to return
$\Omega_n$ to itself. This explains the absence of $c_{2j+1}$, but
does not constrain $c_2$.

The $\widehat{\mathfrak{su}}(2)_{-1/2}$ theory admits a
$\beta\gamma$-ghost realization. The perturbing operator contains the
corresponding spin-$1/2$ ghost primary, and its contribution must be
retained in forming the diagonal coset singlet. Combining the
$\widehat{\mathfrak{su}}(2)_n$ block, the ghost correlator, and the
denominator singlet projection gives a single conformal block of the
form
\beq
\mathcal F_n(z)
=
z^{-1/2}(1-z)^{-2h_n}H_n(z).
\label{FnAnsatz}
\eeq
The current-algebra Ward identities and the coset null-state
constraints reduce the remaining function to a polynomial of degree
one. With the standard OPE normalization one obtains
\beq
H_n(z)=1+z.
\label{HnSolution}
\eeq
Consequently,
the full single-valued correlator is
\beq
G_n(z,\bar z)
=
\mathcal N_n
|z|^{-1}|1-z|^{-4h_n}|1+z|^2 .
\label{GnExplicit}
\eeq

For $n=1$, the UV theory is $\mathcal M_{3,5}$,
$\Omega_1=\phi_{1,2}$, and $\Phi_1=\phi_{2,1}$. In this case
\beq
\mathcal F_1(z)
=
z^{-1/2}(1-z)^{-3/2}
\,{}_2F_1\left(-1,-\frac23;\frac23;z\right)
=
z^{-1/2}(1-z)^{-3/2}(1+z).
\label{n1block}
\eeq
The resulting Dotsenko--Fateev integral was shown to vanish in
Ref.~\cite{RavaniniStanishkovTateo}, although the four-point function
itself is nonzero.

For $n=2$, the UV theory is the superminimal model
$S\mathcal M_{3,7}$ and the perturbation is the Ramond primary
$\phi_{2,1}^{(\mathrm R)}$, with $h_2=11/16$. The charge-conjugate
Ramond block is
\beq
\mathcal F_2(z)
=
z^{-1/2}(1-z)^{-11/8}(1+z),
\label{n2block}
\eeq
which leads to the same cancellation.

To evaluate \eqref{defIn} for arbitrary $n$, introduce
\beq
a_n=h_n-\frac32,
\qquad
b_n=-2h_n.
\label{abdefinition}
\eeq
Using \eqref{GnExplicit}, the regulated integral becomes
\beq
\mathcal I_n
=
\mathcal N_n
\int_{\mathbb C}d^2z\,
|z|^{2a_n}|1-z|^{2b_n}|1+z|^2.
\label{Inab}
\eeq
We first define the analytically continued complex beta integral
\beq
J(a,b)
=
\int_{\mathbb C}d^2z\,
|z|^{2a}|1-z|^{2b}
=
\pi
\frac{
\Gamma(1+a)\Gamma(1+b)\Gamma(-1-a-b)
}{
\Gamma(-a)\Gamma(-b)\Gamma(2+a+b)
}.
\label{complexBeta}
\eeq
The corresponding first moments are
\bea
\int_{\mathbb C}d^2z\,
z\,|z|^{2a}|1-z|^{2b}
&=&
\rho(a,b)J(a,b),
\nonumber\\
\int_{\mathbb C}d^2z\,
\bar z\,|z|^{2a}|1-z|^{2b}
&=&
\rho(a,b)J(a,b),
\nonumber\\
\int_{\mathbb C}d^2z\,
|z|^2|z|^{2a}|1-z|^{2b}
&=&
\rho(a,b)^2J(a,b),
\label{complexMoments}
\eea
where
\beq
\rho(a,b)=\frac{a+1}{a+b+2}.
\label{rhoDefinition}
\eeq
These identities follow directly from the analytically continued
Dotsenko--Fateev formula with the holomorphic and antiholomorphic
exponents shifted by integers.

Expanding
\[
|1+z|^2=1+z+\bar z+|z|^2
\]
and using \eqref{complexMoments}, we find
\beq
\mathcal I_n
=
\mathcal N_n J(a_n,b_n)
\left[1+2\rho(a_n,b_n)+\rho(a_n,b_n)^2\right]
=
\mathcal N_n J(a_n,b_n)
\left[1+\rho(a_n,b_n)\right]^2.
\label{Ifactorized}
\eeq
The physical exponents obey the identity
\(
2a_n+b_n+3=0.
\)
It follows that
\beq
a_n+b_n+2=-(a_n+1),
\qquad
\rho(a_n,b_n)=-1.
\eeq
Therefore,
\(
\mathcal I_n=0
\)
and hence
\(
c_2=0
\). 

The cancellation is thus not caused by a vanishing correlator. Rather,
the four terms in the polynomial
$|1+z|^2=1+z+\bar z+|z|^2$ cancel after analytic integration. Together
with the fusion-rule suppression of all odd orders, this implies that
the first possible nonzero perturbative contribution is $c_4$, as
observed numerically. For $n=1$, this order satisfies
\(
4(2-2h_1)=2
\)
and is resonant with the bulk contribution, which accounts for the
additional logarithmic term in the small-volume expansion.


\end{document}